\documentclass[preprint,aps,prb,amsmath,amssymb,superscriptaddress]{revtex4-2}
\usepackage{graphicx} 
\usepackage{amsmath, amssymb}
\usepackage[left=1.5cm, right=2cm]{geometry}

\usepackage[applemac]{inputenc} 
\usepackage{soul}

\usepackage{natbib}
\usepackage{ulem}
\usepackage{gensymb}

\usepackage{textcomp}
\usepackage[T1]{fontenc}
\usepackage{amsmath}
\usepackage{booktabs}  
\usepackage[caption = false]{subfig}
\usepackage[dvipsnames]{xcolor}
\usepackage{braket}
\usepackage{bm}
\usepackage{dsfont}
\usepackage{adjustbox}
\usepackage{graphicx}
\usepackage{enumitem}
\usepackage{bbm}
\usepackage{tikz}
\usetikzlibrary{arrows.meta,decorations.markings,intersections}

\usetikzlibrary{decorations.markings,intersections}

\tikzset{
  thinarc/.style={draw, line width=0.4pt},
  arrowone/.style={
    postaction={decorate},
    decoration={markings,
      mark=at position 0.50 with {\arrow{Latex[length=2mm]}}
    }
  },
  arrowtwo/.style={
    postaction={decorate},
    decoration={markings,
      mark=at position 0.28 with {\arrow{Latex[length=2mm]}},
      mark=at position 0.72 with {\arrow{Latex[length=2mm]}}
    }
  },
  arrowthree/.style={
    postaction={decorate},
    decoration={markings,
      mark=at position 0.18 with {\arrow{Latex[length=2mm]}},
      mark=at position 0.50 with {\arrow{Latex[length=2mm]}},
      mark=at position 0.82 with {\arrow{Latex[length=2mm]}}
    }
  }
}

\usepackage{physics}
\usepackage{microtype}
\usepackage{txfonts}
\usepackage{cancel}
\usepackage[unicode=true,colorlinks=true,pdfpagemode=UseOutlines,pdfstartview=Fith]{hyperref}

\usepackage{slashed}
\usepackage{orcidlink}

\makeatletter

\usepackage{color}
\usepackage{xcolor}
\usepackage{comment}

\newcommand{\be}{\begin{equation}}
\newcommand{\ee}{\end{equation}}
\newcommand{\bea}{\begin{eqnarray}}
\newcommand{\eea}{\end{eqnarray}}
\newcommand{\beq}{\begin{equation}}
\newcommand{\eeq}{\end{equation}}

\begin{document}
\title{Stoner transitions beyond mean-field in two-dimensional electronic systems: a diagrammatic Monte Carlo study}
\author{Yueh-Chen Lee}
\affiliation{School of Physics and Astronomy and William I. Fine Theoretical Physics Institute, University of
Minnesota, Minneapolis, MN 55455, USA}

\author{Nikolay V. Prokof'ev}
\affiliation{Department of Physics, University of Massachusetts, Amherst, MA 01003, USA}

\author{Andrey V. Chubukov}
\affiliation{School of Physics and Astronomy and William I. Fine Theoretical Physics Institute, University of
Minnesota, Minneapolis, MN 55455, USA}

\date{\today}
\begin{abstract}
Stoner instabilities for fermions with repulsive interaction have been well studied within
the mean-field (ladder) approximation. In this study, we consider two-dimensional
electronic systems with one or two valleys and discuss a Stoner transition beyond the
ladder approximation. At weak coupling, the corrections to the ladder approximation
come predominantly from the renormalization of the particle-hole vertex in the particle-particle channel, and the lowest-order corrections are logarithmically singular in the low-density limit. To investigate the problem beyond the lowest order, we apply
the diagrammatic Monte Carlo algorithm and treat ladder and non-ladder renormalizations
on equal footing. We find that in a one-valley system, a Stoner transition to a
ferromagnetism occurs at low density only if there is a cutoff on the momentum transfer carried by the
interaction. In a two-valley system, the restriction is less severe. Here we find either a
direct Stoner transition into a spin- and valley-polarized state, or a set of two Stoner transitions via an intermediate valley-polarized state. The pathway is controlled by the anisotropy of the fermionic dispersion.
\end{abstract}
\maketitle

\section{Introduction}

In this communication we present the results  of our diagrammatic Monte Carlo (DiagMC) analysis of isospin  Stoner transition~\cite{Stoner1938}  in  two-dimensional (2D) single-valley and multi-valley electronic systems.  The issue attracted substantial interest in recent years because of experimental discoveries of Stoner-type isospin ordering in AlAs quantum wells \cite{Hossain2020,Hossain2021,Hossain2022} and multi-layer graphene systems such as Bernal bilayer graphene (BBG) ~\cite{zhou2022science,holleis2025natphys,Seiler2022}, rhombohedral trilayer ~\cite{Zhou2021,Arp2024,huang2023,blinov2023,Ghazaryan2021,chatterjee2022},  tetralayer~\cite{auerbach2025}  and   
pentalayer~\cite{han2024} graphene, all  in a  displacement field that flattens the bottom of the conduction band and the top of the valence band 
and enhances the density of states (DOS).
From the theory side, Stoner ferromagnetism in a 
single-valley 2D system  with isotropic dispersion  and short-range interaction has been 
  predicted
within mean-field (MF) approximation ~\cite{Zach2024PRBunconventional}.  However, corrections to the MF act against Stoner ferromagnetism~\cite{kanamori1963,irkhin2001,Zach2024PRBlog}.
These corrections are  particularly  relevant in the 
low-density limit  in  2D,  as the irreducible  interaction  in the particle-hole channel
responsible for the Stoner instability within the MF is logarithmically suppressed due to renormalization in the particle-particle channel and becomes too weak to
promote the instability.
This holds even  when the system is tuned to a van Hove singularity~~\cite{ojajarvi2024,kozik2026}. Whether Stoner instability develops at larger densities is not clear.   Earlier variational and diffusion Monte Carlo studies
(VMC and DMC) for  a  system with a parabolic dispersion and unscreened Coulomb interaction found no transition to  ferromagnetism. However, a recent VMC study has found~\cite{Valenti2025} that for a gate-screened Coulomb interaction there exists a range of distances from the gate,  for which a ferromagnetic order does develop.
This last result suggests that screening of the Coulomb interaction is determinative to Stoner ferromagnetism.
Stoner ferromagnetism has also been found in flat-band systems ~\cite{Antebi2024}.
          
In two-valley systems with an electronic structure similar to  AlAs (two elliptical Fermi pockets along $X$ and $Y$ directions in the momentum space),  isospin  Stoner transitions, leading to a valley order and subsequent spin transition to a  simultaneous spin and valley order,  have been detected in VMC calculations~\cite{valenti2024,calvera2025}  for gate-screened Coulomb interaction. A transition to a valley order has also been detected in  the analytical calculations beyond MF ~\cite{Zach2024PRBlog}. 
This again points out to the relevance of gate screening  to  isospin ferromagnetism. 

In our study  we  adopt DiagMC  technique to analyze  how Stoner magnetism depends on the strength of gate screening of the Coulomb interaction.   
A gate-screened Coulomb interaction $U(q)$ is roughly a constant $U_0$ at small momentum transfers  and   decays once $q$ exceeds a certain scale $q^*$.
 We model this by approximating  $U(q)$ by a step function $U(q) = U _0\Theta (q^* -q)$, where $\Theta (x) = 1$ for $x >0$ and $\Theta (x) =0$ for $x <0$.
 
\begin{figure*}
    \centering
    \includegraphics[width=\linewidth]{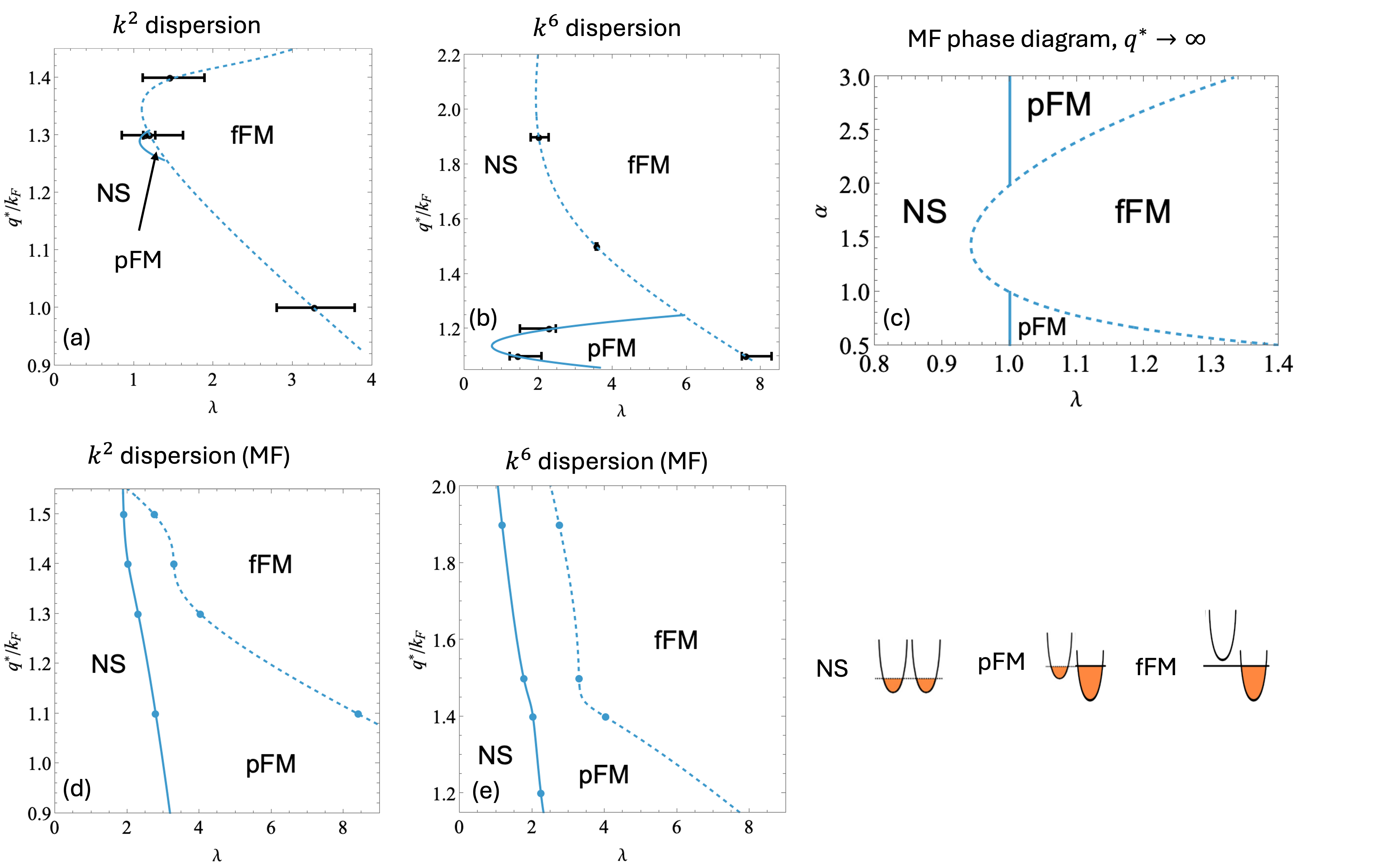}
    \caption{Phase diagrams of single-valley systems with a parabolic dispersion with $k_F=0.1$ (panel (a)) and $k^6$ dispersion with $k_F=0.5$ (panel (b)).  The diagrams are based on the sampling of several values of $q^*$, shown as black dots.  {The corresponding mean-field phase diagrams are shown in  panel (d) and (e), respectively.} Panel (c): The MF phase diagram in units of $\lambda$ and the exponent $\alpha$ of the dispersion  $\epsilon_k=c k^{2\alpha}$ (Ref.\cite{Zach2024PRBunconventional}).
    Solid lines mark a second-order Stoner transition,  to the right of the dashed line 
    the  fully polarized state has the  {lower free energy than the unpolarized state}. Notations:  NS, pFM, fFM are for the normal (unpolarized), partially polarized and fully polarized ferromagnetic state, respectively. We show the schematic depiction of different states at the bottom-right. In a pFM state, the magnetization increases gradually with increasing $\lambda$.}
    \label{fig:phase diagram one valley}
\end{figure*}

We summarize our results for a one-valley system in  Fig.~\ref{fig:phase diagram one valley}. 
 We consider a system of fermions with dispersion relation
  $\epsilon_k=ck^{2\alpha}$, 
where  $c$ has a dimension of energy. 
We measure $k$ in units of the inverse lattice spacing $1/a$.  In these units, $k$ is dimensionless and each component
 of  $k$ is within $-\pi <k_i <\pi$, such that  $k = (k^2_x +k^2_y)^{1/2}$ varies between $0$ and $\pi \sqrt{2}$.   { In our numerical simulations,  we impose a smaller momentum cutoff $k_{max}=1$ to avoid lattice effects.} 
%
 The density of states (DOS) per spin projection  is 
\beq
N(\epsilon) = \frac{1}{4\pi \alpha c^{1/\alpha}} \frac{1}{\epsilon^{(\alpha -1)/\alpha}} 
\eeq
The Fermi energy is $\epsilon_F = c k^{2\alpha}_F$ and 
\beq
N (k_F)  = N_F =\frac{1}{4\pi \alpha c} \frac{1}{k_F^{2(\alpha -1)}} 
\eeq
At $T=0$, the  dimensionless coupling relevant to Stoner transition is 
  $\lambda = U_0 N_F = U_0 \Pi_{ph} (0)$, where $\Pi_{ph} (0)$  is a  static polarization bubble for free fermions in the limit of zero momentum. For our calculations at a finite $T$, we use the same definition of $\lambda$ in terms of $\Pi_{ph} (0)$, but use the value of the  polarization bubble at a finite $T$.  

We find that 
 for a parabolic dispersion, a ferromagnetic transition occurs for $q^*$ comparable to  $k_F$, 
   but not at larger values of $q^*$. 
    {This is distinct from  the mean-field result that  the transition occurs at any value of $q^*$.} 
   The transition is predominantly first-order, except in a narrow window of $q^*$, where it is second order. 
For the $k^6$ dispersion,
a Stoner transition 
persists to larger $q^*/k_F$ due to  the larger DOS $N(\epsilon)$ at small $\epsilon$ (i.e., small $k$) and a  steep 
 variation of the dispersion  {for $k$ close to but still smaller than $k_{max} =1$}, which effectively provides an additional cutoff for the interaction 
  {(e.g. the  energy $\epsilon_k$ changes between $0.1c$  and $c$  for   $k$ between   0.68 and 1)}. The transition is 
 first-order for larger $q^*$ and second-order for smaller $q^*$.   
  Below  the second-order transition,  the  system remains in a 
  partially polarized  state in a wide range of couplings.
 In the calculations we used $k_F = 0.1$ for $k^2$  dispersion and $k_F =0.5$ for $k^6$ dispersion. 

\begin{figure*}
\centering
\includegraphics[width=\linewidth]{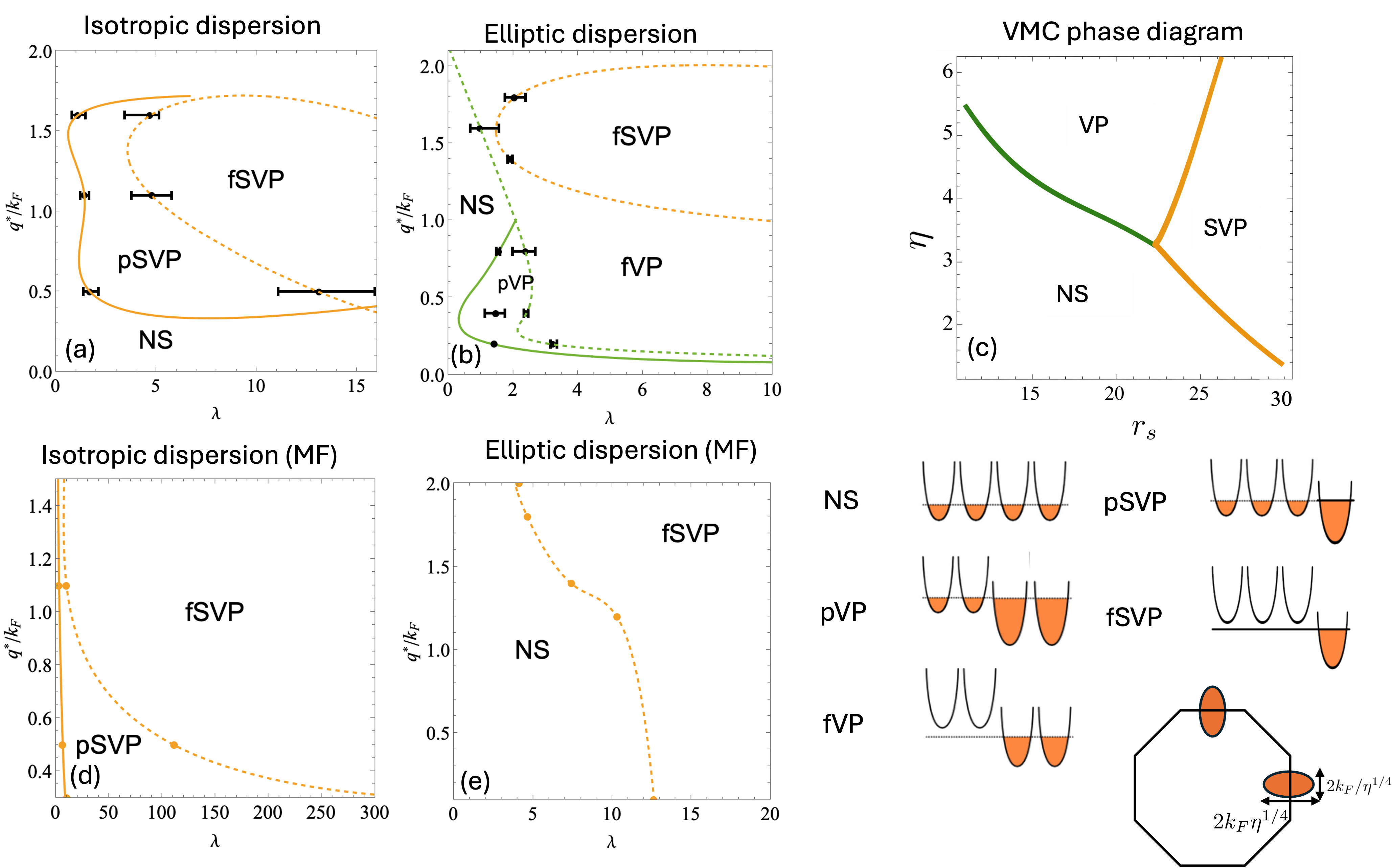}
    \caption{ Phase diagram of a two-valley system with an isotropic dispersion relation  (panel (a) ),  and elliptic dispersion with the  anisotropy factor $\eta=5.2$ (see Eq.~\ref{eq:elliptic}) 
    (panel (b)).  {The corresponding mean-field phase diagrams are shown in panels (d) and (e), respectively.} 
    The symmetry-breaking phases are valley polarized (VP) and spin-valley polarized (SVP). The symbol p/f means the system is partially/fully polarized. The color code represents the type of broken symmetry 
     below the transition: valley (green) and both (orange). 
    Panel (c):  The variational quantum Monte Carlo phase diagram adapted from Ref. ~\cite{calvera2025} 
     and shown here for comparison with DiagMC data. The diagram is 
    for a system with a gated Coulomb interaction,  as a function of $\eta$ and the dimensionless Wigner-Seitz radius $r_s=\frac{me^2}{4\pi\varepsilon\hbar^2}\frac{1}{\sqrt{\pi n}}$, where $\varepsilon$ is the dielectric constant of the medium, and $n$ is the density of fermions.
    Under this panel  we show the schematic depiction of various different states and elliptic  Fermi surfaces (bottom right).
    }
    \label{fig:phase diagram two valley}
\end{figure*}

We summarize our results for  a two-valley system in Fig.~\ref{fig:phase diagram two valley}.  
Here,  {we placed the centra of the $X$ and $Y$ valleys at $(\pi,0)$ and $(0,\pi)$ of the Brillouin zone. 
We keep the  momentum cutoff at the same $k_{max}=1$ as in a one-valley system. We note that 
$k_{max}$ is smaller than half of the distance between the two pockets, $\pi/\sqrt2$ so that the dispersions are not mixed in the calculations.}
 We consider the same Hubbard 
 interaction with a cutoff at $q^*$ and a parabolic, yet  elliptic dispersion for each valley. We label the valleys  by $\gamma = \pm 1$:  {$\epsilon_{\mathbf k, \gamma =1} = c(\sqrt\eta k^2_x + (1/\sqrt\eta) k^2_y)$,  $\epsilon_{\mathbf k, \gamma =-1} = c(\sqrt\eta k^2_y + (1/\sqrt\eta) k^2_x)$.  }
 We use the same definition of the dimensionless coupling as before,  $\lambda = U_0 \Pi_{ph} (0)$ and set with $k_F=0.1$. 

For an isotropic dispersion ($\eta =1$)  spin and valley symmetries are
either unbroken or broken simultaneously~\cite{valenti2024,calvera2025,Zach2024PRBlog}. For this case we find 
valley/spin ordering transition for $q^*$ comparable to $k_F$. The transition is second-order into partially  polarized valley/spin state. A full polarization is then reached at larger couplings. 
For elliptic dispersion,  we find at  a larger $q^*$  a sequence  of two first-order transitions: first into a valley-polarized state and then into a valley and spin polarized state. At smaller $q^*$, we find  a   second-order valley transition, while the spin symmetry is never broken. 
 {We note that this is different from the mean-field approximation in which 
 spin and valley orders are degenerate and become simultaneously broken for all values of $q^*$}

There is another  interesting aspect of  the analytical studies of an isospin Stoner transition in 2D:  both one-valley and two-valley systems with $k^2$ and $k^4$  dispersion were shown within MF  to be at the boundary between a fist-order and second-order Stoner transition~\cite{Zach2024PRBunconventional,Zach2024PRBlog}. For these systems,  the  MF  analysis 
for the interaction without a cutoff 
shows that an  isospin susceptibility diverges upon approaching the  critical coupling, as is expected for a second-order transition,   yet 
 at the transition the free energy is flat as a function of isospin magnetization, and 
immediately  below the transition  the magnetization jumps to its larger possible value,  totally depleting 
 fermions from one or two Fermi surfaces. 
 In our DiagMC study of the model with the cutoff for the interaction at $q^*$ 
 we did find a discontinuous transition accompanied by the divergence of susceptibility at the special values of $q^*$, where Stoner transition changes from second to first-order.  {However, we do not claim that the 
  free energy is flat at the transition point as we only compare the energies of the unpolarized and the fully polarized states without probing intermediate values of the order parameter.
   On general grounds, we do not expect the free energy in our case to be completely flat at the critical point. 
    Depending on the shape of the free energy at intermediate values of the order parameter, there is either  a transition to a maximally polarized state right when the susceptibility diverges, or a transition into a partially-ordered state. We also note in this regard that we did numerical calculations at finite temperature, when the free energy profile is not flat even for the interaction with no cutoff (Ref.~\cite{megan}). }

 The rest of the paper is structured as follows. In Sec.~\ref{sec:model} we introduce the models and the DiagMC algorithm, the details of which are covered in App.~\ref{app:DiagMC}. We present the results for one- and two-valley systems in Sec.~\ref{sec:one_valley} and ~\ref{sec:two valley}, respectively. We conclude with Sec.~\ref{sec:summary}.


\section{Models and methods}\label{sec:model}

For a one-valley system, we consider a model of two-dimensional fermions 
with the effective Hamiltonian $H'=H-\mu N$ in the form 
\begin{equation}
    H'=\sum_{\mathbf{k},\sigma}\left(\epsilon_{\mathbf{k},\sigma}-\mu\right)c^\dag_{\mathbf{k},\sigma} c_{\mathbf{k},\sigma}+\sum_{\mathbf{k},\mathbf{p},\mathbf{q},\sigma,\sigma'}\frac{U(q)}{2}c^\dag_{\mathbf{p}-\mathbf{q},\sigma} c^{\dag}_{\mathbf{k}+\mathbf{q},\sigma'}c_{\mathbf{k},\sigma'} c_{\mathbf{p},\sigma},
\end{equation}
where $\sigma$ is the spin index,  $\mu$ is chemical potential in the unpolarized state and    $U(q)$  is a dual-gated Coulomb interaction.
As we said above, we assume that the dispersion $\epsilon_k = c k^{2\alpha}$ and the potential 
\begin{equation}
    U(q) = \left\{\begin{aligned}
        &U_0, \,q<q^*\\
        &0,\, \textrm{otherwise,}
    \end{aligned}
    \right.
    \label{eq:U}
\end{equation}
 where  $q^* \sim 1/d$ and  $d$ is the distance to the screening gate, which can be varied in the  experiments.
In the limit $q^*\rightarrow\infty$, $U(q)$ is reduced to a contact Hubbard interaction  $U=U_0$.
   
A ferromagnetic order parameter is  a vector: 
\begin{equation}
    {\vec \Delta}=(1/N) \sum_k \langle c^\dagger_{k, i} {\vec \sigma}_{i,j} c_{k,j} \rangle,
\end{equation}
where ${\vec \sigma}_{i,j} = (\sigma^x_{i,j},   \sigma^y_{i,j}, \sigma^z_{i,j})$, and $  \sigma^l_{ij}$ are $2 \times 2$ 
Pauli matrices. 
 {For simplicity, here and later in the paper, we 
set 
the Bohr magneton $\mu_B$ to be $1$.}

For a two-valley system, we consider the low-energy model with two elliptic electron pockets  centered  at the $X$  and $Y$ points in  the Brillouin zone,  hereafter labeled by the valley indices $\gamma=\pm1$,
respectively~\cite{shayegan2006,Hossain2020,valenti2024,calvera2025}  (see Fig.~\ref{fig:phase diagram two valley} bottom right). 
 The free part of the Hamiltonian is 
\begin{equation}
H'_0=\sum_{\mathbf{k},\sigma,\gamma}\left(\epsilon_{\mathbf{k},\sigma,\gamma}-\mu\right)c^\dag_{\mathbf{k},\sigma,\gamma} c_{\mathbf{k},\sigma,\gamma}
\end{equation}
with the dispersion relation given by
\begin{eqnarray}
    \epsilon_{\mathbf{k},\gamma=1}=c \left(\sqrt\eta k_x^2+\frac{1}{\sqrt\eta} k_y^2\right)\\
    \epsilon_{\mathbf{k},\gamma=-1}= c \left(\frac{1}{\sqrt\eta} k_x^2+\sqrt\eta k_y^2\right),
    \label{eq:elliptic}
\end{eqnarray}
where $\mathbf{k}$ is the deviation from the $X(Y)$ point of the Brillouin zone for $\gamma=1(-1)$ and $\eta$ is the  measure of the anisotropy of the dispersion ($\eta=1$ for isotropic dispersion). The  model is invariant under $\eta\rightarrow1/\eta$ and  below we only consider $\eta\geq1$.  The interacting Hamiltonian  for this system  contains  intra-valley and inter-valley density-density interactions. In both, the momentum transfter is between fermions from the same valley, hence  both are $U(q)$: 
\begin{eqnarray}
    H'_{\textrm{int}}&&=\sum_{\mathbf{k},\mathbf{p},\mathbf{q},\sigma,\sigma',\gamma,\gamma'}\frac{U(q)}{2}c^\dag_{\mathbf{p}-\mathbf{q},\sigma,\gamma} c^{\dag}_{\mathbf{k}+\mathbf{q},\sigma',\gamma'}c_{\mathbf{k},\sigma',\gamma'} c_{\mathbf{p},\sigma,\gamma},
\end{eqnarray}
 We again approximate $U(q)$ by Eq.  (\ref{eq:U}).   We follow earlier works \cite{Zach2024PRBlog,valenti2024,calvera2025} and neglect  the exchange interaction between the two valleys
  as this interaction is at a  large momentum roughly equal  to the distance between the centers of $X$ and $Y$ pockets.

A generic order parameter for a two-valley system  has both scalar and vector components and can be schematically expressed as 
\begin{equation}
    \Delta= (1/N) \sum_k \langle c^\dagger_{k, i,l} T_{ij}S_{l,m} c_{k, j,m} \rangle,
\end{equation}
where $T=\delta,\vec{\gamma}$ and $S=\delta,\vec{\sigma}$,  $\delta$ is an  identity matrix and $\vec {\gamma}$ and  $\vec{\sigma}$ are Pauli matrices in the valley and spin space, respectively. 
The  components of this order parameter are  valley-polarization (VP, $T=\gamma_z,S=\delta$), ferromagnetism (FM, $T=\delta,S=\vec{\sigma}$), staggered ferromagnetism (SFM, $T={\gamma}_z,S=\vec{\sigma}$), real/imaginary spin density wave (SDW, $T=\gamma_{x/y},S=\vec{\sigma}$) and charge-density-wave (CDW, $T=\gamma_{x/y},S=\delta$). 
It is pointed out in ~\cite{Zach2024PRBlog,valenti2024,calvera2025} that for  elliptical  Fermi surfaces, the susceptibilities for density-wave orders 
   are smaller than  for VP, FM and SFM. For this reason, below we 
   will only discuss VP and  FM/SFM.
   
   We report below the results of two calculations.  In the first, we compute  the order  parameter susceptibilities as functions of dimensionless coupling $\lambda=U_0\Pi_{ph}(0)$ for different $q^*$, where $\Pi_{ph}(0)$ is the bare susceptibility  {per fermion flavor}. These  calculations are particularly relevant for the cases when the transition is second-order as in this situation an order parameter susceptibility diverges at the transition. 
   In the second calculation, we compute the free energy of a fully polarized spin or valley or spin and valley  ordered state and compare it with the  free energy of the unpolarized state.   These calculations are mostly relevant for the cases when the Stoner transition is  first-order. Since we do not have the free energy profile for each value of $U$ and $q^*$ as a function of the order parameters, this provides an upper bound of the coupling below which the first-order transition occurs. In principle, it is possible that the system remains partially polarized until larger $\lambda$. However, it is pointed out in Ref.~\cite{megan} that even at finite temperature, the system with Hubbard interaction exhibits a partially polarized state only in a fairly narrow window of $U_0$ before it is fully polarized in the case of a first-order transition. We will use the analysis of the free energy also for the cases of  a second-order transition to estimate when the system becomes fully polarized.  For obvious reasons, this estimate yields the lower bound of the coupling at which  the system becomes fully polarized.

\subsection{Evaluation of the  susceptibility}

In DiagMC, we work in the imaginary time  basis to avoid dealing with the pole structure of the Green's function. The  bare Green's function of a fermion with  spin $\sigma$  and valley $\gamma$ is
\begin{equation}
    G_{\sigma,\gamma}(\mathbf{k}, \tau)=\left\{
    \begin{aligned}
        -e^{-(\epsilon_\mathbf{\textbf{k},\sigma,\gamma}-\mu_\sigma)\tau}(1-n_F(\epsilon_{\mathbf{k},\sigma,\gamma})),\,\tau>0,\\
        e^{-(\epsilon_\mathbf{\textbf{k},\sigma,\gamma}-\mu_\sigma)\tau}n_F(\epsilon_{\mathbf{k},\sigma,\gamma}),\,\tau\leq0,
    \end{aligned}
    \right.
    \label{eq:Green's function}
\end{equation}
where $n_F(\epsilon_\mathbf{\textbf{k},\sigma,\gamma})$ is the Fermi function for spin $\sigma$ and valley $\gamma$.  

In our calculation, we  evaluate the fully dressed 4-fermion interaction vertex and neglect the renormalizations of the quadiparticle dispersion, the chemical potential, and the quasiparticle residue. 
Some of these renormalizations can be 
 incorporated into re-definitions of the bare fermionic propagator and of  $U_0$, as it is assumed in a Fermi liquid theory, but in the diagrammatic calculations one has to add counter-terms to avoid multiple-counting. 
  {For an estimate, we calculated the quasiparticle residue $Z$ and mass renormalization to the leading order
  in $\lambda$. For $Z$, the first nonzero contribution comes from the second-order diagram shown in Fig.~\ref{fig:Z}. For one-valley system with isotropic parabolic dispersion and the  cutoff at  $q^*=k_F$, we  obtain $Z\approx0.72$ at the critical value of the coupling $\lambda=1$. The mass renormalization comes already from the first-order Fock diagram. We find $m^*/m\approx1.37$ at the critical coupling. The correction to the chemical potential is less essential, as the dressed $\mu$ rather than the bare one is related to the density. We believe that the fact that the leading renormalizations of $Z$ and $m^*/m$ are relatively weak justifies our approach, in which we neglected these renormalizations. Indeed, very near the second-order transition lines, $m^*/m$ diverges and $Z$ vanishes at the smallest frequencies. However, this generally has little effect on the transition lines as they are determined by frequency-integrated $m^*/m$ and $Z$. 
 \begin{figure}
     \centering
     \includegraphics[width=0.5\linewidth]{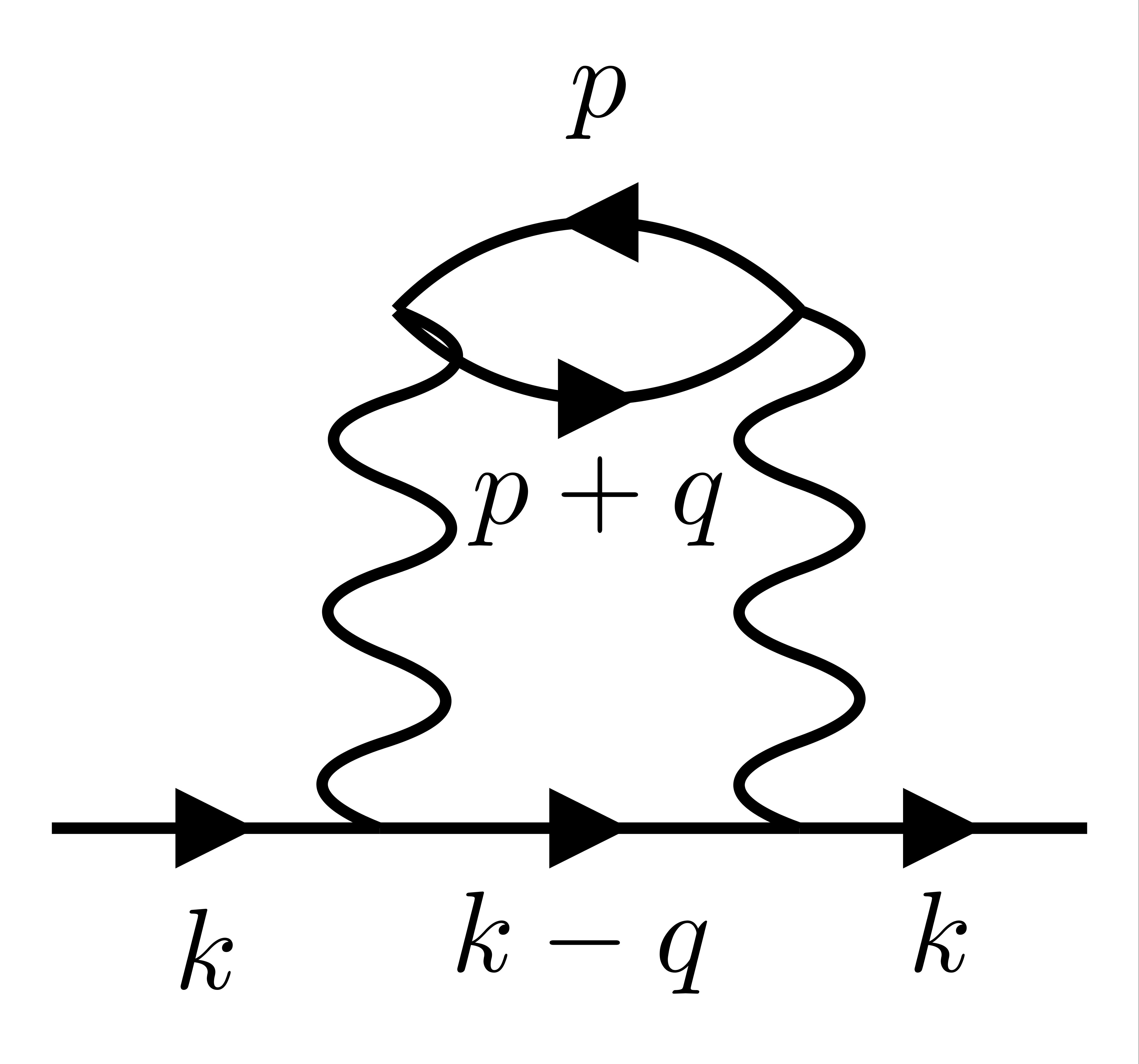}
     \caption{The second-order diagram that contributes to the quasiparticle residue $Z$.}
     \label{fig:Z}
 \end{figure}
 }

\begin{figure}
    \includegraphics[height=0.8\textheight]{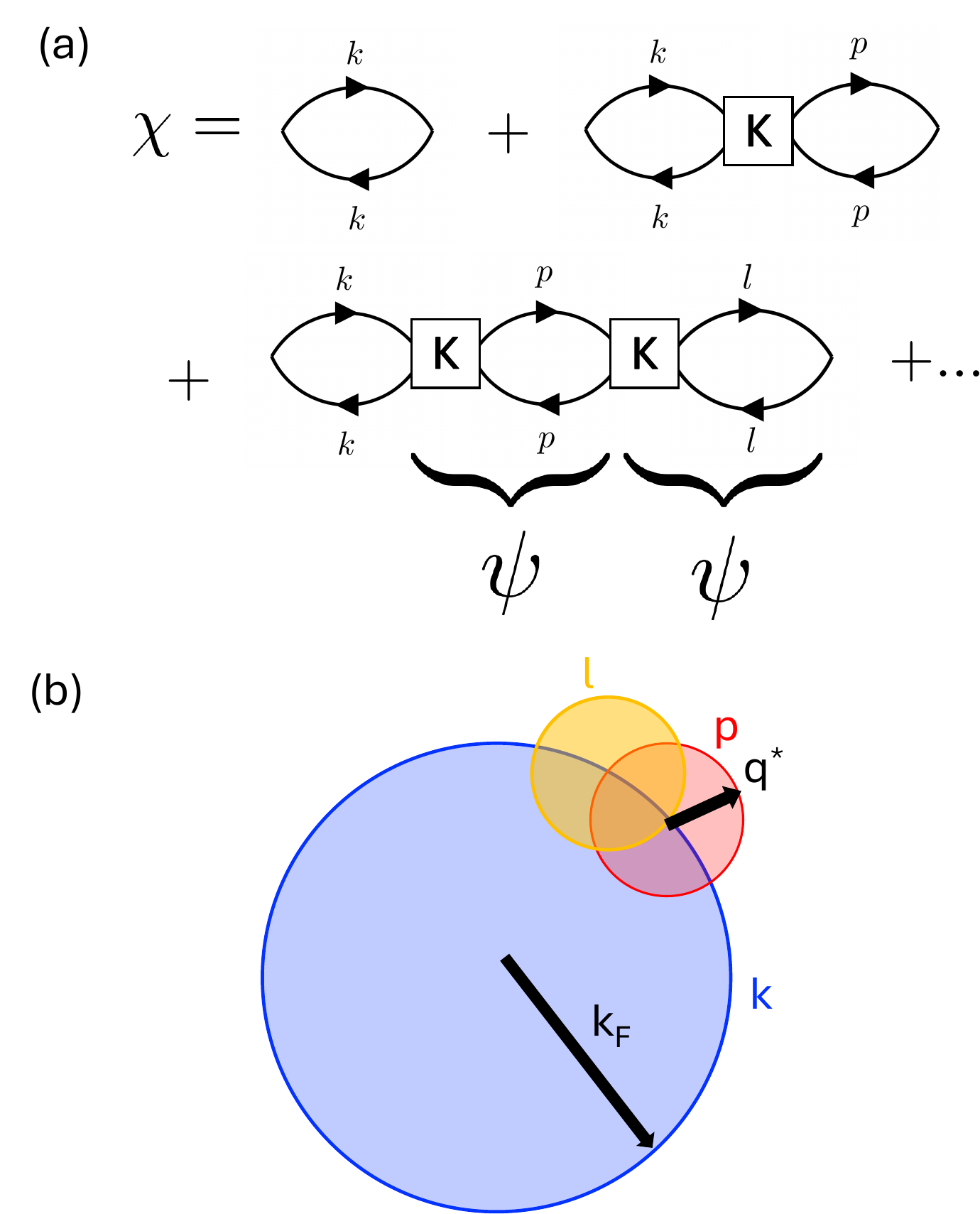}
    
    \caption{(a) Summation of the ladder diagrams. The second term is what we define as $\chi_{2PI}$.
     The quantity $\psi$ plays the same role as $l=0$ spin Landau parameter  in Fermi liquid theory.
      (b) The Fermi surface and the range of integration restricted by the cutoff $q^*$ with a circular Fermi surface. The range of integration over $\textbf{k},\textbf{p},\textbf{l}$ in (a) are marked in blue, red and yellow, respectively.}
    \label{fig:cutoff}
\end{figure}

Our first  objective is to calculate the full static susceptibility, which can be represented as an infinite diagrammatic series  in Fig.
 ~\ref{fig:cutoff}(a).   The first term is the bare particle-hole bubble with zero external momentum 
 $\chi_0^s =$ 
  { $n_f\Pi_{ph}(0)$, where $n_f$ is the number of fermion flavors, spin times valleys.} 
 The momentum integral in this term
  is confined  to the Fermi surface,  and all Fermi points contribute (the blue circle in Fig.  ~\ref{fig:cutoff}(b)). 
  We evaluate this term analytically using  a regularization by a finite temperature. The second term can be written  as  {
\begin{widetext}
\begin{equation}
    \chi^s_{2PI}  
    \equiv\int_{\tau,\tau_1,...,\tau_4} 
    \int_{\mathbf{k},\mathbf{p}} \mathbf{\sigma}_{ik}G_{i,\gamma} (\mathbf{k},\tau_1)G_{k,\gamma}(\mathbf{k},\tau_2)
   K(\mathbf{k},\mathbf{p},\tau_1+\tau_2,\tau_3 +\tau_4)\delta_{ij}\delta_{kl}
  G_{j,\gamma}(\mathbf{p},\tau_3)G_{l,\gamma}(\mathbf{p},\tau_4)\sigma_{lj},
\label{tt_5}
\end{equation}
\end{widetext}}
where the summation over repeated spin indices is assumed, and we used the short-hand notation $\int_{\mathbf{k}}\equiv\int\frac{d^2k}{(2\pi)^2}$ and  $\int_{\tau_i} \equiv \int_0^{1/T} d \tau_i$. We used finite $T$ for regularization of the contribution from double poles. 
  The variable  $K(\mathbf{k},\mathbf{p},\tau_1+\tau_2,\tau_3+\tau_4)$ is the fully dressed irreducible  four-point function that includes all  diagrams which  do not contain cross-sections with two  fermionic lines with the same momentum and frequency. 
  The spin dependence of $K = K_{\alpha,\gamma; \beta,  \delta} $ in general contains two 
 combinations of spin indices: $K = K_1 \delta_{\alpha \beta}\delta_{\gamma \delta} + K_2 \delta_{\alpha \gamma}\delta_{\beta \delta}$. We kept  only the first combination in (\ref{tt_5}) because the $K_2$ part vanishes
  after a convolution with the side $\sigma$-functions. 
 Because the $K_2$ part vanishes identically, we don't need to single out $K_1$ part in the diagrammatic 
   series for $K$.

In the MF approximation,  $K$ is the same as the  bare interaction $U(q)$. 
The momentum integration in Eq.  (\ref{tt_5}) over $\textbf{k}$ and $\textbf{p}$ is confined to 
 momenta infinitesimally close to the Fermi surface.   One integration,  say   $\int _\textbf{k}$,   is over 
 the entire Fermi surface (the blue circle in Fig. \ref{fig:cutoff}(b)).   The second integral  over $\textbf{p}$  is, however,  
 restricted to an arc of the Fermi surface,  enclosed by the circle centered at $\textbf{k}$ with radius $q^*$,  due to the cutoff in  the interaction. Since the Fermi surface is circular, the integral over $\textbf{p}$ in the restrictive region is independent of where $\textbf{k}$ is located on the Fermi surface. We define it as the restrictive bubble with zero external momentum as $\tilde{\Pi}_{ph}(0)$.   
 \footnote{This restriction is relevant for $q^* < 2k_F$. For larger $q^*$, 
the region, over which $\textbf p$ is integrated, covers the entire Fermi surface, and 
 $\tilde\Pi_{ph}(0)$ is the same as $\Pi_{ph}(0)$.  }
  The same argument holds for integration over $\textbf{l}$ in the third term of the ladder series in  Fig. \ref{fig:cutoff}(a)
   and for all  higher-order terms.  
     The full susceptibility for $K = U_0$  can be expressed  as a geometric series:
\begin{equation}
 {
    \chi^s=\frac{n_f\Pi_{ph}(0)}{1-U_0\tilde\Pi_{ph}(0)}.}
    \label{eq:ladder}
\end{equation}
The Stoner criterion 
for 
the onset of an isospin ferromagnetism is 
$U_0\tilde\Pi_{ph}(0)=1$. This is similar to a conventional criterion for a Hubbard interaction, but now $\tilde\Pi_{ph}(0)$ is smaller than $\Pi_{ph}(0)$ due to the cutoff that we imposed on the momentum transfers in the ladder series ,
hence the Stoner condition is met at a higher $U_0$. 

\begin{figure}
      \centering
      \includegraphics[width=\linewidth]{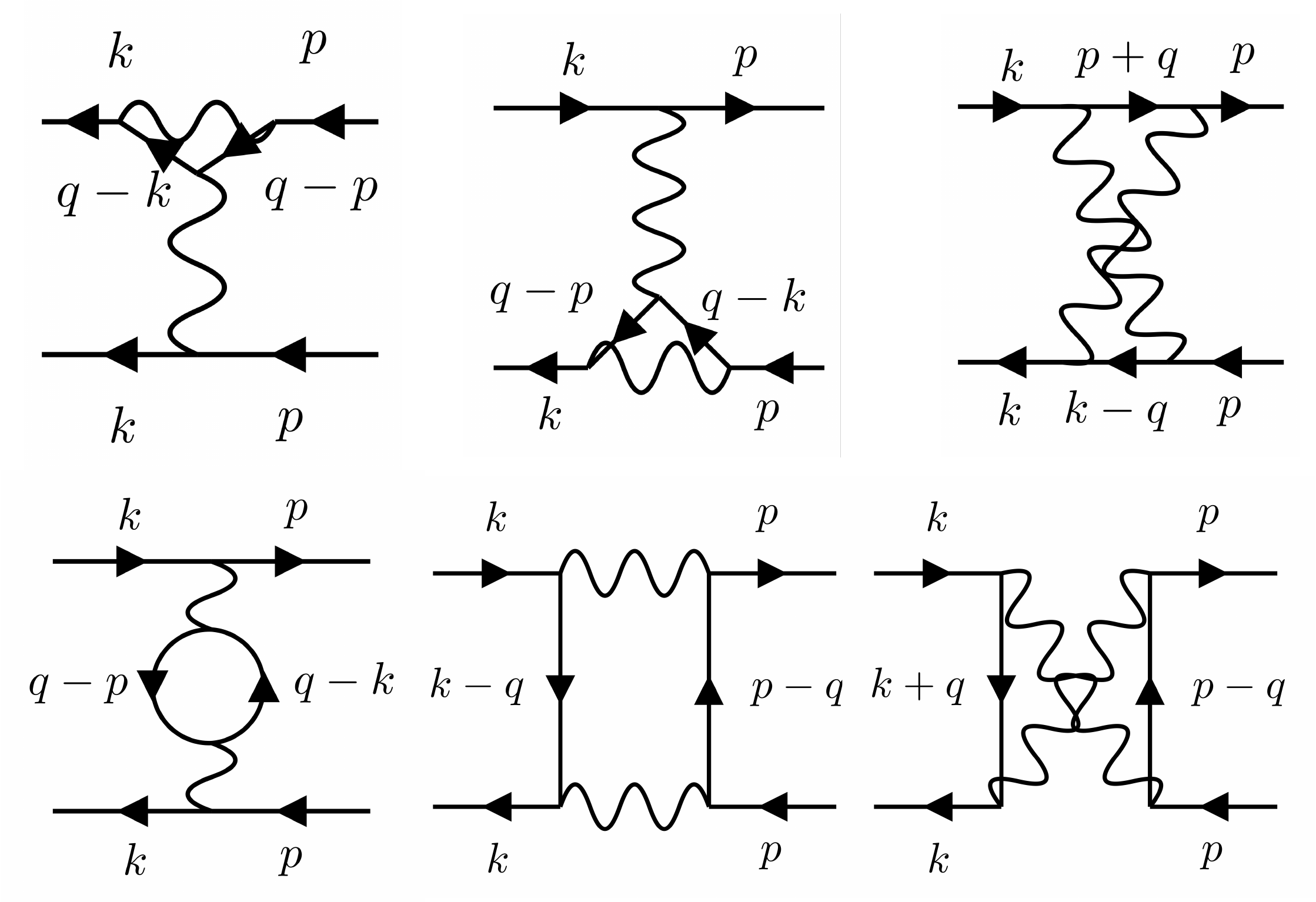}
      \caption{The two-particle irreducible four point function $K$ to second order in $U_0$.  The last two Aslamazov-Larkin diagrams  do not contribute to spin susceptibility but contribute to valley susceptibility for a non-isotropic, elliptic dispersion \cite{Zach2024PRBlog}.} 
      \label{fig:second order}
  \end{figure}

We now go beyond mean-field and include the contributions to $K$  from 
 the diagrams which do not contain particle-hole bubbles with zero frequency transfer and vanishingly small momentum transfer. We  show these diagrams  at order $U^2_0$ in  Fig.~\ref{fig:second order}.
We compute $\chi^s_{2PI}$, obtain the ratio 
\begin{equation}
    \psi^s = \frac{\chi^s_{2PI}}{\chi^s_0},
    \label{eq:psi}
\end{equation}
 and express the full $\chi_s$ as
\begin{equation}
    \chi^s\approx\frac{\chi^s_0}{1-\psi^s},
\label{tt_7}
\end{equation}
The condition  for the onset of  ferromagnetism  from the divergence of the  spin susceptibility is  then $ \psi^s=1$.

\begin{figure}
     \centering
     \includegraphics[width=0.5\linewidth]{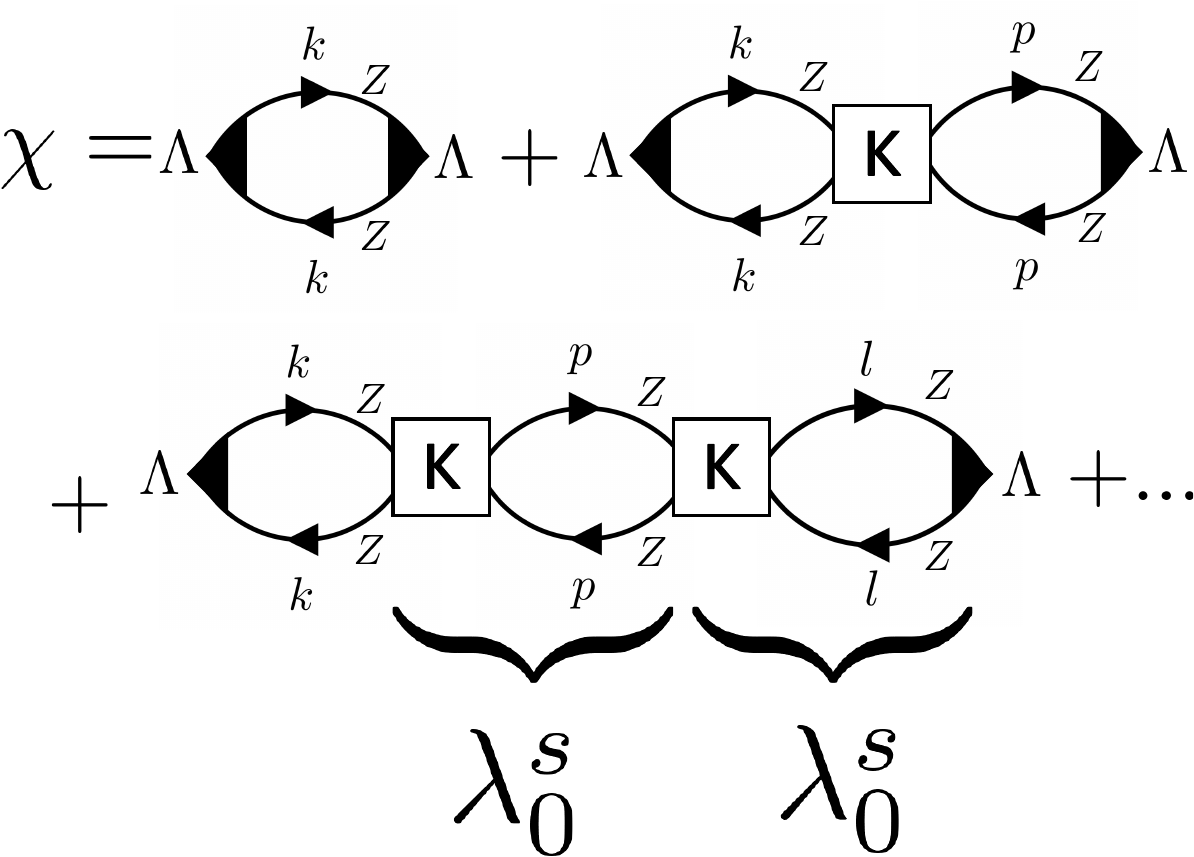}
     \caption{Ladder diagrams for the spin susceptibility in a Fermi-liquid theory. The black-shaded area is the side vertex $\Lambda$, which includes renormalizations from fermions located away from the Fermi surface. Every Green's function contains a quasiparticle residue $Z$. 
     }
     \label{fig:bubble_FL}
 \end{figure}

Our reasoning here is similar to the one used to calculate  the full $\chi^s$ in a Fermi liquid theory  (see e.g., \cite{Chubukov_2018a}). 
 The diagrammatic series for the spin susceptibility in a  generic Fermi liquid are shown in  Fig. \ref{fig:bubble_FL}.
   The Green's functions are  propagators of fermions near the Fermi surface $G = Z/(\omega -
    (k-k_F) k_F/m^*$,  where $m^*$ is the renormalized mass.     The fully dressed side vertices in the  bubble series are labeled as $\Lambda$. 
   Both $Z$ and $\Lambda$  are renormalized by  fermions located away from the Fermi surface  and for this reason  are input parameters  for Fermi liquid theory.
    By Ward identity, $Z \Lambda =1$, hence  $\Lambda$ and $Z$  compensate each other. 
    A single bubble in a Fermi liquid theory  then has almost the same form  as for free fermions:  $\chi^s_0 = 2 \mu^2_B (m^*/m)  \Pi_{ph} (0)$.
      The  two-loop susceptibility  $\chi^s_{2PI}$ in a Fermi liquid  is  expressed as the product of $((m^*/m)\Pi_{ph} (0))^2$, $Z^2$   and the  quasiparticle vertex function $\Gamma^\omega (\mathbf{k},\mathbf{p})$.  The $\sigma$ matrices in the side vertices select the  zero partial component  of $\Gamma^\omega (\mathbf{k},\mathbf{p})$. The product of this component,  $(m^*/m) \Pi_{ph} (0)$ and $Z^2$  is known as $l=0$ spin Landau parameter $\lambda^s_{0}$.   Aside from $Z^2$ and $m^*/m$, the Landau parameter  is given by the same    Eq. \ref{eq:psi}  as our $\psi^s$. The full susceptibility  in a Fermi liquid theory is given by the same equation as   (\ref{tt_7}) with $\lambda^s_0$ instead of $\psi^s$ and extra $m^*/m$ in the overall factor.   From this perspective,  our  computation  of $\chi^s$ parallels the one  in a Fermi liquid theory. We caution, however, that 
      the Landau parameter $\lambda^s_0$ contains the product $Z^2 m^*/m$, which by  itself depends on the interaction $U(q)$.    For this reason, the Landau parameter  is  not simply proportional to the  quasiparticle vertex  function (i.e., to $K$ in our case), and is not guaranteed to keep increasing when $U(q)$ increases.   As we said, we neglect this complication an conjecture that we  capture Stoner physics even if we neglect fermionic $Z$-factor  and mass renormalization. 


In our calculation, we  evaluate  $\chi^s_{2PI}$ up to sixth order in $U_0$ numerically following the DiagMC algorithm introduced in Appendix.~\ref{app:update} and extrapolate to infinite order using the resummation technique which we outline in Appendix.~\ref{app:resummation}. To ensure the irreducibility, we only select the diagrams with no two Green's functions carrying the same momentum (except those connected to the spin vertices, as we are calculating the order with zero momentum).

In the two-valley system, we  calculate the valley susceptibility $\chi^{v}$ in the same manner as $\chi^s$. The only difference is that we put the Pauli matrices in the valley space $\gamma_z$ instead of in the spin space $\vec{\sigma}$. 

\subsection{Evaluation of the free energy}

The free energy $F$ consists of the kinetic and interacting parts. For a single isospin with given density $n$, the kinetic part is calculated in the standard manner:
\begin{equation}
    F_K(n)=\mu n-T\int_\mathbf{k}\log\left[1+e^{(\mu-\epsilon_\mathbf{k})/T}\right],
    \label{Fkin}
\end{equation}
where the chemical potential $\mu(n)$ is solved self-consistently using particle number conservation. Suppose there are $n_0$ particles for each isospin in the unpolarized state, then the difference of the kinetic part of the free energy between the unpolarized and fully  polarized ferromagnetic state in a one-valley system is given by
\begin{equation}
    \Delta F_K=F_K(2n_0)-2F_K(n_0).
\end{equation}
In the two-valley system with isotropic dispersion, when a transition occurs, the spin and valley symmetries are simultaneously broken, and the system directly enters a spin-valley polarized (SVP) state, where there are one major carrier and three minor carriers. In this case, we compare the kinetic energy difference between the fully polarized SVP  state (fSVP) and the unpolarized state
\begin{equation}
    \Delta F_K=F_K(4n_0)-4F_K(n_0),
\end{equation}
where $n_0 $ here denotes the fermion density per isospin. In the system with elliptic dispersion, the system can undergo a cascade of transitions, where spin and valley symmetries are broken at different interaction strengths $U_0$, and there exists an intermediate spin or valley polarized phase, in which either one of the symmetries is broken. In this case, we calculate the kinetic energy difference between either the fully ferromagnetic/valley polarized state (fFM/fVP) and the fSVP phase
\begin{equation}
    \Delta F_K=F_K(4n_0)-2F_K(2n_0).
\end{equation}
On the other hand, the interaction part of the free energy $F_U$ corresponds to the Luttinger-Ward (LW) functional, which is the sum of closed-loop diagrams with no external vertices. For a single-valley system within MF approximation, we only take the lowest order diagrams in $U_0$, i.e., Hartree and Fock diagrams. They give
\begin{eqnarray}
    F_U^{\textrm{Hartree}}&&=\frac{U(0)}{2}\left(\sum_{\sigma,\gamma}\int_{\mathbf{k}}G_{\sigma,\gamma}(\mathbf{k},0)\right)^2=2U_0n_0^2\\
    F_U^{\textrm{Fock}}&&=-\frac{1}{2}\sum_{\sigma,\gamma}\int_{\mathbf{k}}\int_{\mathbf{q}}G_{\sigma,\gamma}(\mathbf{k},0)\,U(q)\,G_{\sigma,\gamma}(\mathbf{k}+\mathbf{q},0)\\
    &&=-\frac{1}{2}\sum_\sigma\int_{\mathbf{k}}\int_{\mathbf{q}}n_F(\epsilon_{\mathbf{k},\sigma,\gamma})\,U(q)\,n_F(\epsilon_{\mathbf{k}+\mathbf{q},\sigma,\gamma}).
\end{eqnarray}
There is no double counting here as we keep fermionic Green's function as bare, with no renormalizations.
For the Hubbard interaction we have $F_U^{\textrm{Fock}}=-1/2F_U^{\textrm{Hartree}}$ in the unpolarized state and $F_U^{\textrm{Fock}}=-F_U^{\textrm{Hartree}}$ in the fFM state.
 Within this approximation,  Stoner transition in a system with $\epsilon_k \propto k^{2\alpha}$ is first order when $1\leq  \alpha \leq2$ and second order for $\alpha >2$ and $\alpha <1$. 
To go beyond MF, we again employ DiagMC to calculate $F$
up to 6th order and extrapolate to infinite order in the same was as for the susceptibility. 

\section{One-valley system}\label{sec:one_valley}

\subsection{Parabolic dispersion}
\begin{figure*}
    \centering
    \includegraphics[width=\linewidth]{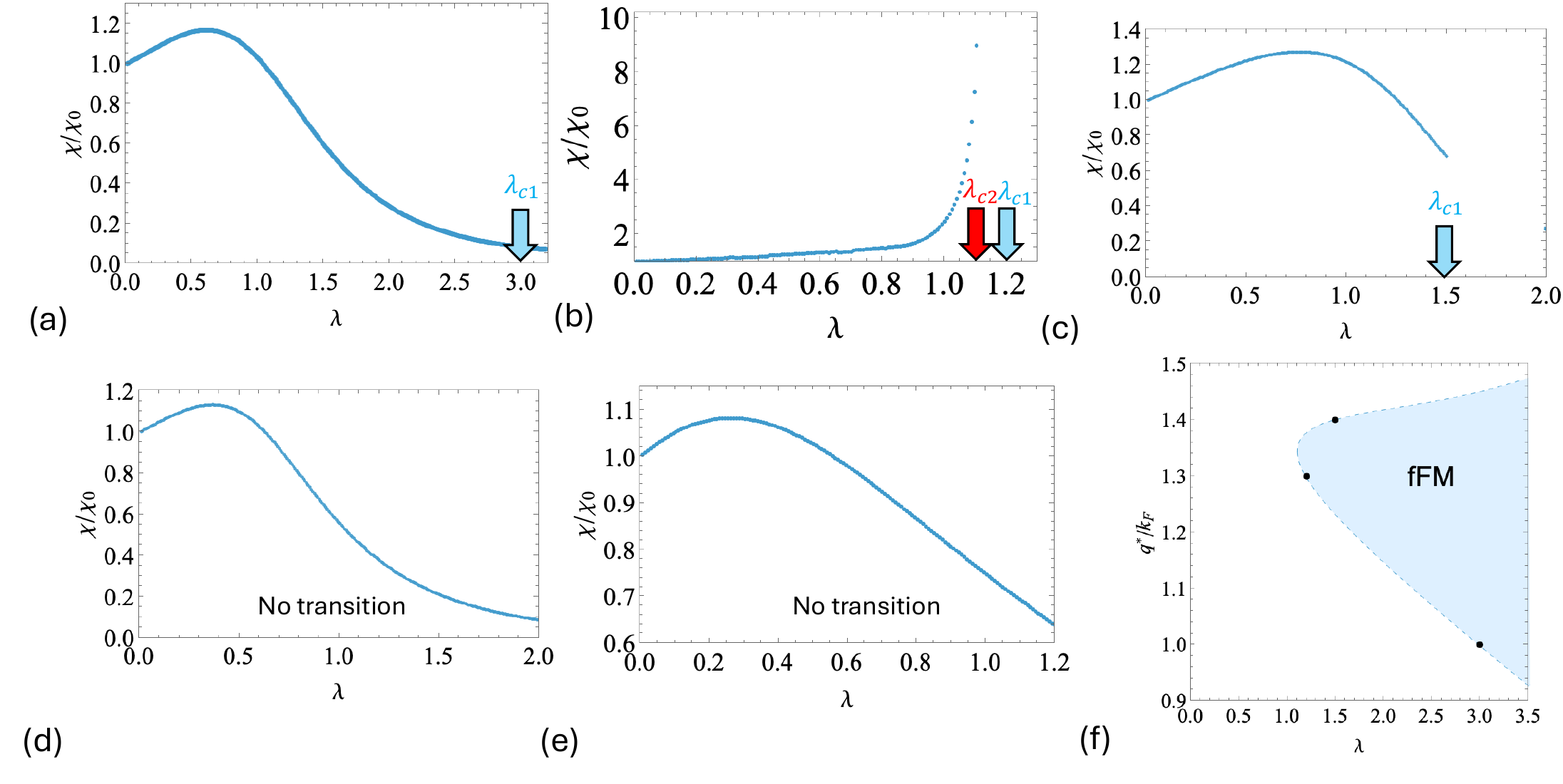}
    \caption{(a)-(e) Full susceptibility in units of its bare value for a single Fermi surface with isotropic parabolic dispersion and $q^*=1,1.3,1.4,1.5k_F$ and  $q^*\rightarrow\infty$. The blue arrows mark when the free energy of the fully polarized state becomes lower than that of the unpolarized 
    state  and the red arrows mark when the susceptibility diverges. (f) The parameter range where the
     lowest energy state is a fully polarized ferromagnet.  The black dots are the values of $q^*$ which 
     we sampled.}
    \label{fig:parabolic}
\end{figure*}

\begin{figure*}
    \centering
    \includegraphics[width=\linewidth]{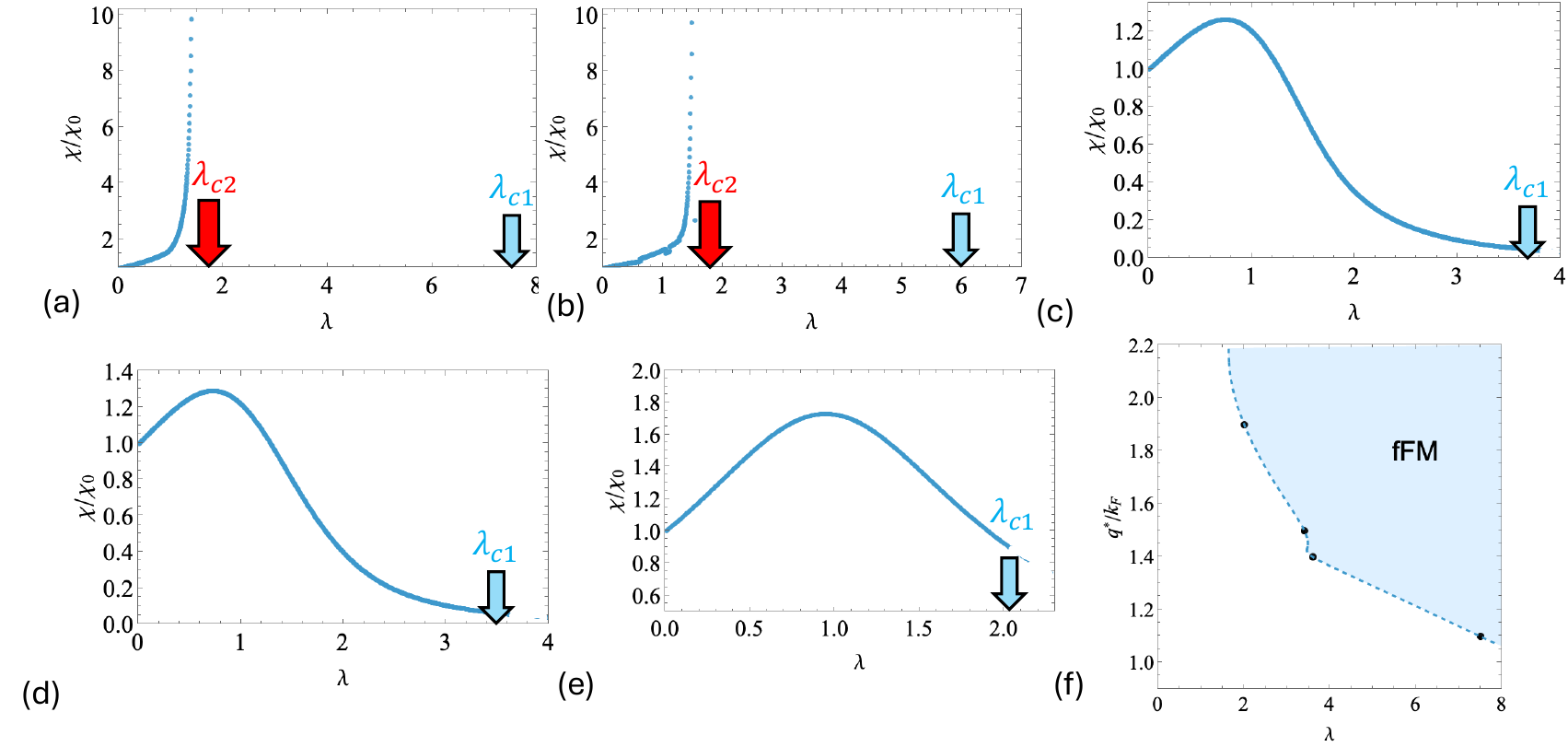}
    \caption{(a)-(e) Full susceptibility in units of its bare value for a single circular Fermi surface with $k^6$ dispersion. Panels (a)-(e) are for 
     and $q^*/k_F =1.1,1.2,1.4,1.5,1.9$.  The blue arrows mark when the free energy of the fully polarized state becomes lower than that of the unpolarized 
    state  and the red arrows mark when the susceptibility diverges. 
       (f) The parameter range where the
     lowest energy state is a fully polarized ferromagnet.  The black dots are the values of $q^*$ which 
     we sampled.}
    \label{fig:k6}
\end{figure*}
\begin{figure*}
    \centering
    \includegraphics[width=\linewidth]{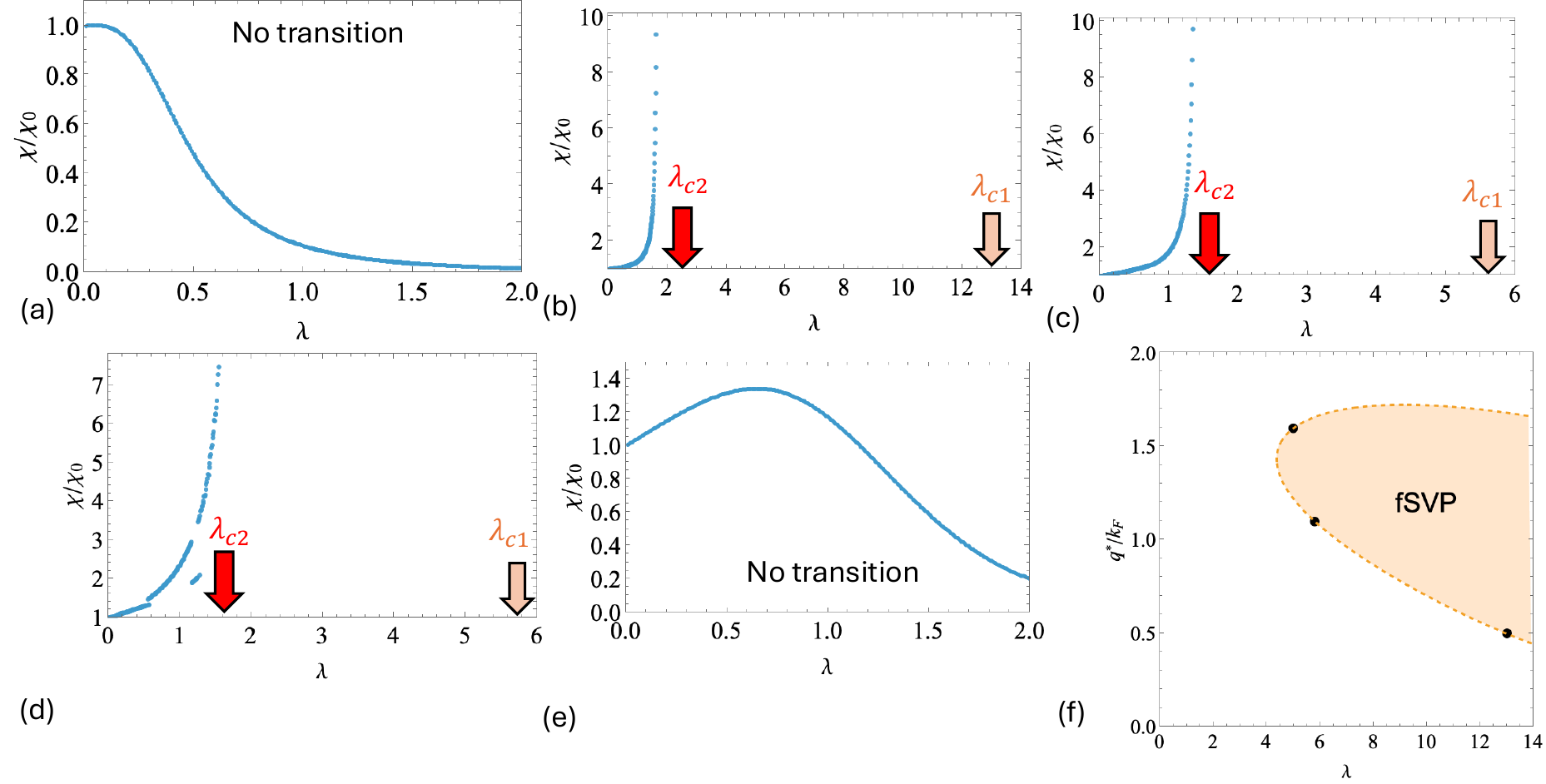}
    \caption{(a) - (e) Full susceptibility for a two valley system with isotropic dispersion with $q^*=0.3, 0.5, 1.1, 1.6, 1.8k_F$. The orange arrows mark when the free energy of the fully polarized state becomes lower than that of the unpolarized 
    state, and the red arrows mark when the susceptibilities in both spin and valley channels diverge.  (f) The parameter range where the
     lowest energy state is spin- and valley-polarized (fSVP).  The black dots are the values of $q^*$ which 
     we sampled.}
    \label{fig:two valley circular}
\end{figure*}
\begin{figure*}
    \centering
    \includegraphics[width=\linewidth]{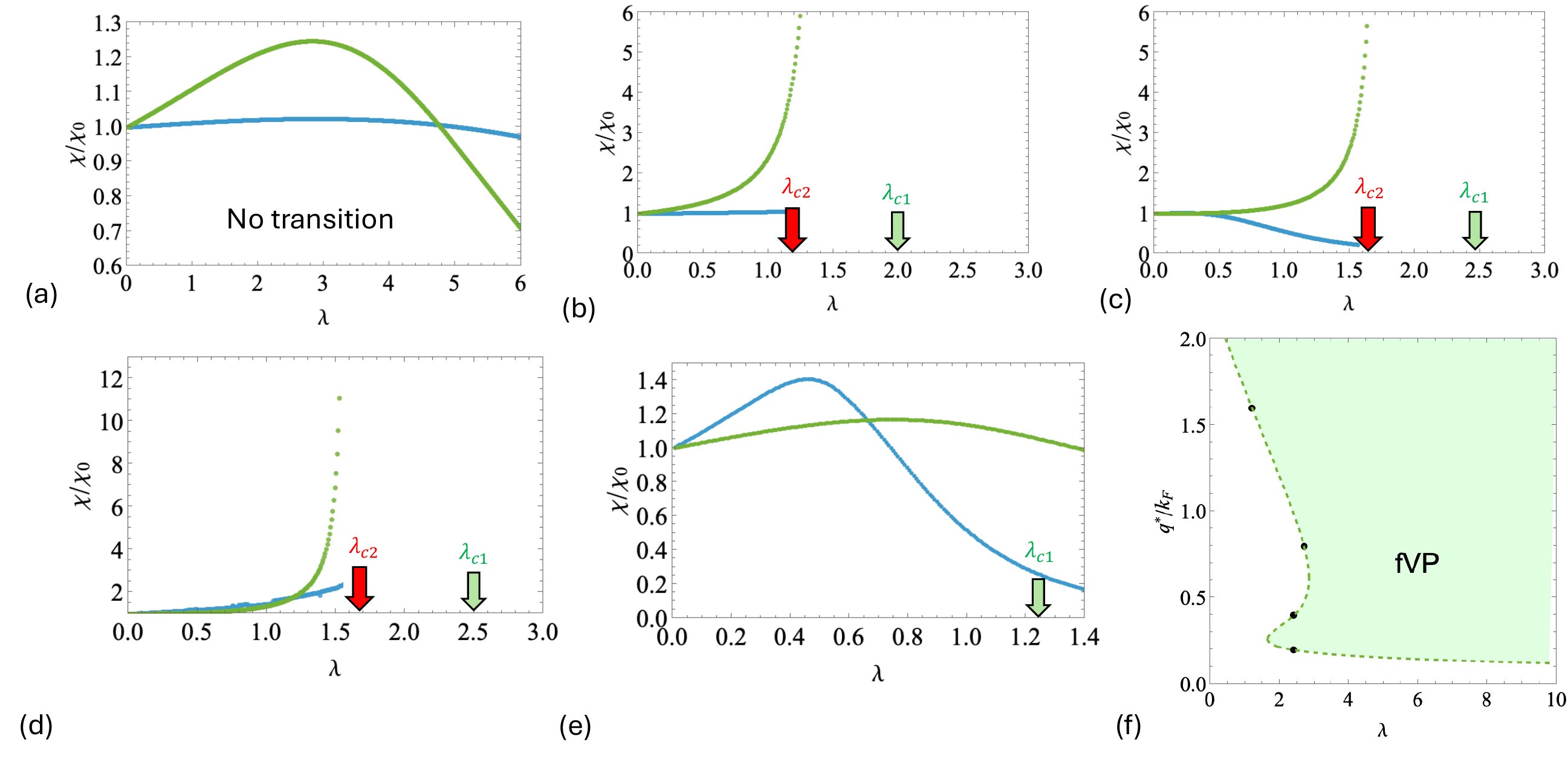}
    \caption{(a) - (e) Full susceptibility  for a two valley system with elliptic dispersion with $\eta=5.2$ and  $q^*=0.1, 0.2, 0.4, 0.8, 1.6k_F$. The blue curves are the  susceptibility in the FM/SFM channel, and the green curves are the susceptibility in the VP channel. The green arrows mark where the system  becomes fully valley-polarized, and the red arrows mark when the valley susceptibility diverges (the spin susceptibility does not diverge in all the cases). (f) The parameter range where the
     lowest energy state is fully valley-polarized (fVP).  The black dots are the values of $q^*$ which 
     we sampled.}
    \label{fig:elliptic}
\end{figure*}

We first  consider the case where $\epsilon_k= c{k^2}$.
 For practical  reasons we set   $T=0.25\epsilon_F$.  This is not a  truly low $T$, but we expect to capture the key physics already at this $T$, even though the precise location of the Stoner transition may be somewhat different from that at $T=0$.  We recall that in the MF approximation, the Stoner transition 
 occurs at $\lambda=1$. Below  we label as $\lambda_{c2}$ the critical coupling strength, at which the susceptibility diverges, and as $\lambda_{c1}$  the coupling strength above which it  becomes energetically more favorable for the system to be fully polarized than 
   unpolarized.  We note that $\lambda_{c1}$, defined this way, is generally smaller than the 
   coupling at which the system becomes fully polarized simply because the energy of  a partially polarized state can be smaller than that of the fully polarized state.  This, however, is not an issue for us as our key goal is to locate the Stoner transition and detect whether it is first or second order. 

We  show the behavior of the susceptibility for various $q^*$ in   Fig.~\ref{fig:parabolic}(a)-(e)
 (values of $q^*$ are in the figure caption). In panel (f) of this figure we  highlight the region where the system is in the fully polarized ferromagnetic state (fFM). We already presented the  full phase diagram  in Fig.~\ref{fig:phase diagram one valley}(a).

At smaller $q^* \leq  k_F$, the unpolarized state has the lowest energy at $\lambda < \lambda_{c1} \approx 3$ 
 (the blue arrow in Fig.~\ref{fig:parabolic}) and the fully polarized 
   state has the lowest energy at larger $\lambda > \lambda_{c1}$.
 At somewhat larger $q^*$  there is a narrow range at $q^* \approx 1.3k_F$, where the system undergoes a 
 second-order transition at $\lambda = \lambda_{c2} \geq 1$.  Outside this range, the transition remains first order. At $q^*\geq 1.5k_F$, neither first- nor second-order transition is observed. 
  {This is distinct from the mean-field approximation, in which the transition occurs at all values of $q^*$ (Fig.~\ref{fig:phase diagram one valley}(d)).
 }

\subsection{$k^6$ dispersion}

We recall that in the MF approximation,  a system with  dispersion $\epsilon_k=ck^{6}$  undergoes a second-order Stoner transition  (see Fig.~\ref{fig:phase diagram one valley}(c)).
  We show the  results of our  beyond MF calculations in Fig.~\ref{fig:k6}. One can see from the top panel that for  $q^* \leq 1.2k_F$,   the system undergoes a second-order transition aat $\lambda_{c2} \leq 2$ 
  (panels (a) and (b)).    
From the analysis of the free energy we find another, larger $\lambda_{c1}$, where  the  fully polarized state becomes energetically more favorable than the unpolarized state. It is natural to assume that in 
the interval $\lambda_{c2} <\lambda < \lambda_{c1}$,  the system is in a partially polarized state.
 
 For  $q^*$ larger than $1.2k_F$, the susceptibility does not diverge, but the fully polarized state still becomes energetically more favorable at $\lambda > \lambda_{c1}$.  We  indicate this in Figs.~\ref{fig:k6}(c)-(e)
  by blue arrows at $\lambda = \lambda_{c1}$. For these $q^*$, which extend  up to $q^* \leq 2 k_F$, the system undergoes the first-order transition into a fully polarized state. We show the range of the fully polarized state in Fig.~\ref{fig:k6}(f) and  show the full phase diagram in Fig.~\ref{fig:phase diagram one valley}(b).
   Note that $\lambda_{c1}$ decreases as $q^*$ increases. We did not study $q^*>2k_F$;  it is possible that for these $q^*$,  $\lambda_{c1}$ gets close to 1. The  extended ranges of $q^*$ where either first- or second-order transitions occur  for two complementary reasons:  the DOS gets larger at small $k$, where the dispersion is flat, and the steep dispersion at  {$k$ close to 1} effectively imposes an additional cutoff on the interaction reducing the strength of the renormalization acting against Stoner instability.

\section{Two-valley system}\label{sec:two valley}

\subsection{Isotropic dispersion relation}

As pointed out in ~\cite{Zach2024PRBlog}, in the case of isotropic dispersion, the spin and valley symmetries are simultaneously broken when a transition occurs. We therefore only calculate the energy difference between the unpolarized state and the fSVP state. We also calculate the susceptibility in the spin channel to probe the existence of a second-order transition. 

We present the results of our calculation of  susceptibilities for  different $q^*$ 
   in
Fig.~\ref{fig:two valley circular}(a)-(e). In panel 
   (f)
of this figure we highlighted the range 
    where 
the system is fully spin- and valley-polarized.  The  full phase diagram is presented  in Fig.~\ref{fig:phase diagram two valley}(a). 
At $q^* > 1.8k_F$, there is no transition of either first- or second-order.   At $q^*=1.6k_F$, the system exhibits a second-order transition at $\lambda_{c2}$ slightly smaller than 2, whereas the free energy of the fSVP state is lower than that of the unpolarized state at $\lambda_{c1}\approx6$. 
Compared to a single-valley system with parabolic dispersion, a  first-order transition occurs at a higher $q^*$ in a two-valley system.
This suggests that the two-valley system shows a higher tolerance to  interactions  with  large-momentum transfer.   For $q^*$ in  the range  $0.5 k_F < q^*<1.6k_F$,
   the system exhibits a second-order transition at $\lambda_{c2} \leq 2$ and becomes  fully polarized at $\lambda_{c1}$ ranging from $5$ to $14$, depending on $q^*$. At $q^*=0.3k_F$, the divergence of susceptibility disappears again, and the system again shows no transition at all.

\subsection{Elliptic dispersion relation}

It is shown in ~\cite{Zach2024PRBlog} that in a system with Fermi surface anisotropy $\eta$ and inter-valley density-density interaction the logarithmic renormalization  from the particle-particle channel is suppressed by $\frac{2\sqrt\eta}{\eta+1}$,  whereas  for  intra-valley interaction is not suppressed. 
Because inter-valley interaction favors VP order, 
    it is expected that this order will 
    start to dominate over FM/SFM. 
%
%
Furthermore,   for large enough $\eta$,  critical $\lambda_c$  for the onset of VP can be even smaller than   that in the MF.  

In panels (a)-(e) of Fig.~\ref{fig:elliptic} we show the behavior of susceptibilities as functions of  $\lambda$ for   different $q^*$ at $\eta =5.2$,   and in panel (f) we show the range where the system favors  full valley polarization. We  did not  detect the  transition to a ferromagnetic state. The full phase diagram is shown in Fig.~\ref{fig:phase diagram two valley}(b). We see that the system with elliptic dispersion undergoes a second-order transition into the VP state
 for $ 0.5k_F < q^* <k_F$ and first-order  transition  for larger $q^*$ (up to $q^* \approx 2k_F$ that we analyzed) .  In the range where the transition is second-order, 
 the valley susceptibility $\chi^{\textrm{VP}}$ diverges at  $\lambda_{c2}$ varying between 1 and 2, while {$\lambda_{c1}\approx$} 2 to 3. 
   {This is different from the mean-field result in Fig.~\ref{fig:phase diagram two valley}(e), in which spin and valley orders are degenerate and are thus both orders develop at the same coupling.
  }
 \begin{figure*}
    \centering
    \includegraphics[width=\linewidth]{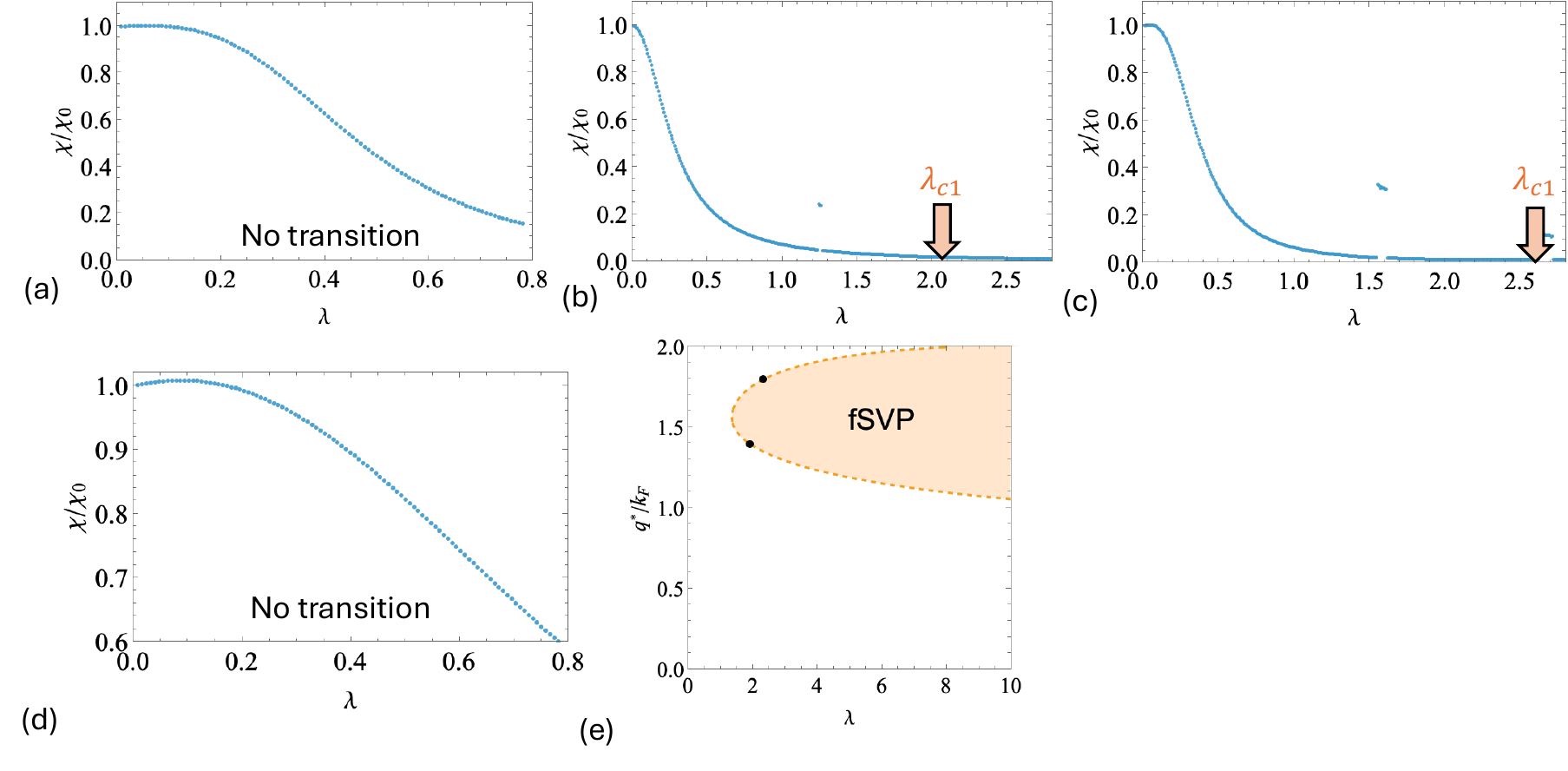}
    \caption{Full spin susceptibility for a transition from VP to SVP state with $\eta=5.2$ and $q^*=1.2,1.4,1.8,2.0k_F$. The orange arrows show where the system  becomes fully spin- and valley-polarized. Note that here $k_F$ refers to the Fermi momentum of the unpolarized state instead of the VP state.}
    \label{fig:SVP_out_of_VP}
\end{figure*}
\begin{figure*}
    \centering
    \includegraphics[width=\linewidth]{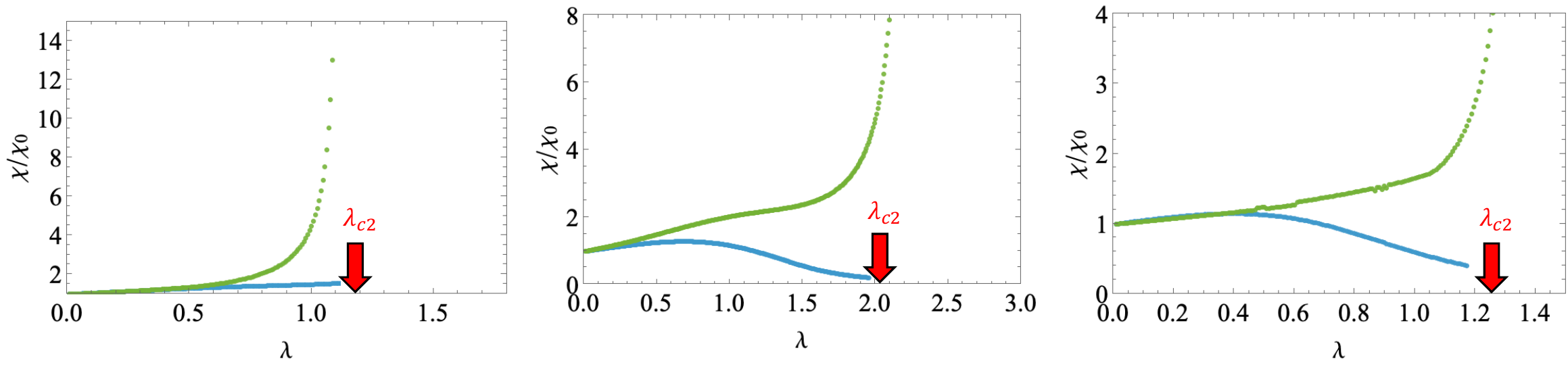}
    \caption{Ferromagnetic (blue) and valley (green)   susceptibilities   for $q^*=k_F$ and $\eta=2.2, 3.6, 4.8$ (left to right). The red arrows mark when the valley susceptibility diverges . The  spin susceptibility does not diverge in all the cases. }
    \label{fig:eta}
\end{figure*}
The easiest way to see why the valley polarization transition in systems  with  elliptic dispersion  is present  at a smaller $q^*$ than in systems  with isotropic dispersion is to calculate the kinetic energy  that the interaction has to  overcome to obtain the valley polarization.   For a system with two spins and two valleys, the density  of a component with a given spin projection $\sigma$ and valley index  $\tau$ can be parameterized by the three order parameters $\zeta_1,\zeta_2,\zeta_3$, where $\zeta_1$ measures the FM order, while $\zeta_2$ and $\zeta_3$ measure  VP and SFM orders, respectively. The relation is 
 ~\cite{Zach2024PRBunconventional}
\begin{equation}
    n_{\sigma\gamma}=n_0(1+\sigma\zeta_1+\gamma\zeta_2+\sigma\gamma\zeta_3),
\end{equation}
subject to the constraint $-1\leq\zeta_i\leq1$ and $n_{\sigma\gamma}\geq0$, where $n_0$ is the fermion density per isospin in the unpolarized state.   For a parabolic dispersion relation $\epsilon_k=k^2$, the  kinetic energy is given by
\begin{equation}
    E_K={8\pi n_0^2}(1+\zeta_1^2+\zeta_2^2+\zeta_3^2).
\end{equation}
For the fully polarized VP state (fVP) , $\zeta_2=\pm1$ and $\zeta_1=\zeta_3=0$. On the other hand, for  the  fully polarized fSVP state,  which develops  for isotropic dispersion,  $\zeta_1^2=\zeta_2^2=\zeta_3^2=1$. We see  that the kinetic energy increase that the interaction has to overcome  is three times larger for the fSVP state than for the  fVP state and speculate that this is the reason why  for elliptical Fermi surfaces the VP transition persists at smaller $q^*$.

In ~\cite{valenti2024,calvera2025}, it is suggested that at an even stronger interaction, there exists a second transition from VP to SVP, in which the spin symmetry is also broken. In Fig.~\ref{fig:SVP_out_of_VP} we show the spin susceptibility in the VP state and the range where the fSVP state is lower than the fVP state. It can be seen that the system exhibits a first-order transition at the range $1.4k_F\leq q^*\leq1.8k_F$, where the transition point $\lambda_{c1}$ is between $2$ and $3$. Outside of this range, there is no transition at all. The spin susceptibility, on the other hand, shows no sign of divergence throughout the range of $q^*$ we sample.

Looking at the behavior of susceptibilities, to connect the picture with the system of isotropic dispersion, where all the orders are degenerate and $\chi^{\textrm{FM/SFM}}$ and $\chi^{\textrm{VP}}$ diverge simultaneously, it is worth comparing the susceptibilities for different Fermi surface anisotropies $\eta$. In Fig.~\ref{fig:eta}, we show  $\chi^{\textrm{FM/SFM}}$ and $\chi^{\textrm{VP}}$ at $q^*=k_F$ for different values of $\eta$. One can see that at $\eta=2.2$, $\chi^{\textrm{VP}}$ diverges at a $\lambda_{c2}$ slightly larger than one, whereas $\chi^{\textrm{FM/SFM}}$ increases, but only by a very small amount. At larger anisotropies $\eta=3.6$ and $4.8$, $\chi^{\textrm{VP}}$ still diverges, but $\chi^{\textrm{FM/SFM}}$ shows no sign of enhancement at all. This confirms that $\chi^{\textrm{FM/SFM}}$ tends to increase at smaller anisotropies, albeit it is normally accompanied by the divergence of $\chi^{\textrm{VP}}$ at a smaller $\lambda$.

Since the transition breaks spin-rotation symmetry out of a single-valley system, it is worth comparing with the ferromagnetic transition as in Sec.~\ref{sec:one_valley}, and one notices three major differences: 1. The transition begins to occur at a larger $q^*$ than that in the single-valley case. 2. The transition is always first-order when it occurs. To explain these differences, one first notices that $q^*$ is in units of $k_F$ in the unpolarized state. However, in the fully VP state, the area of the Fermi surfaces is actually doubled in area since all the particles are in the same valley. Therefore, the threshold of $q^*$ to remove the logarithmic divergence in the particle-particle channel is larger. For the same reason, the reduced bubble $\tilde{\Pi}_{ph}(0)$ is integrated over a smaller portion of the Fermi surface, suppressing the enhancement of the susceptibility.

\section{Summary}\label{sec:summary}
In this communication, we study the Stoner-type transition in both one- and two-valley systems in two dimensions. We introduce a simplified interaction with a cutoff on momentum transfer to model the gate-screened Coulomb interaction. We show that the existence of the transition and its order depend on a number of factors: In one-valley systems with parabolic dispersion, one needs to set the cutoff in the interaction to be of the order of the Fermi momentum to see a transition to a ferromagnetic state, whereas in the $k^6$ case the transition remains at higher cutoffs. In the latter case, one can tune the transition from first- to second-order by decreasing the cutoff. Compared to the case of parabolic dispersion, 
    a dispersion, which is flatter at small momentum and steeper at high momentum,  relaxes the condition to larger cutoffs. 
In two-valley systems, we show that, compared to the system with isotropic dispersion, which can directly enter a spin and valley polarized state, a system with elliptic dispersion prefers to be valley polarized over ferromagnetic. By tuning the cutoff, the transition can either be first- or second-order. In the case of elliptic dispersion, upon increasing interaction, the valley-polarized state is followed by a subsequent ferromagnetic transition and becomes both spin and valley-polarized at large enough cutoffs, where the transition is first-order, whereas at small cutoffs the spin transition does not occur.

\section{Acknowledgments}
The authors would like to thank the fruitful discussions with E. Kozik, K. Van Houcke, F. Werner, V. Calvera, Y. Yu, R. D. Mayrhofer, E. K.  Kokkinis, A. Rastogi, K. R. Islam and B. Currie. YL  wants to thank the assistance from V. Dantas Meireles and the training from Y.-J. Kao and K.-H. Wu in running numerical simulations.   The research by YL and  AVC   was supported by  the National Science Foundation grant NSF:DMR-2325357.      AVC and NVP acknowledge support from the Simons Foundation grant SFI-MPS-NFS-00006741-07 for the Simons Collaboration on New Frontiers in Superconductivity.

\appendix
\section{Technical details of Diagrammatic Monte Carlo algorithm}\label{app:DiagMC}
\subsection{Updates and measurement}\label{app:update}
Suppose we want to calculate an observable $\mathcal{O}$ perturbatively using the diagrammatic expansion. We can write down the general expression as
\begin{equation}
\mathcal{O}=\sum_{n=0}^\infty\sum_{T_n}\int \mathcal{D}_{T_n}(x_1,\dots ,x_n)dx_1dx_2\dots dx_n,
\end{equation}
    where the  sum is the order of the expansion, $n$ and  over diagrams of different topologies at a given $n$,  $T_n$, and  the integral is over  internal variables $x_1\dots x_n$. In our case, $x_i$ are momenta and imaginary times, and we also have to sum up over  isospin labels of the Green's functions, $G$.  
%
     This expression is then sampled using  
     Markov chain Monte Carlo updates with acceptance 
     ratios based on the detailed balance: 
if we have  a configuration $A$, characterized by a set of internal variables $\{x_i\}$, and we want to perform an update to change it to the configuration $B$, then the acceptance probability is given by 
\begin{equation}
    P_{\textrm{accept}}=\frac{P(A\rightarrow B)W(B)}{P(B\rightarrow A)W(A)},
\end{equation}
where $P(A\rightarrow B)$ and $P(B\rightarrow A)$ are the probabilities of selecting the updates $A\rightarrow B$
     and $B\rightarrow A$, 
     and $W(A)$ and $W(B)$ are the statistical weights of configurations involved in the update 
     (they are moduli of the products of $G$ and $U$ functions). 
The configuration space of Feynman diagrams contributing to $\mathcal{O}$ is then sampled
      with probability  density $\propto |\mathcal{D}_{T_n}(x_1,\dots ,x_n)|$     
      ~\cite{VANHOUCKE201095}, and the 
 Monte Carlo estimate for $\mathcal{O}$ is the expectation value of $\text{sgn}(\mathcal{D}_{T_n})$ (the normalization 
is provided by the exact analytic evaluation of the lowest-order contribution).
Since $\mathcal{D}_n=a_nU_0^n$, 
    we solve for $a_n$ by considering $U_0=1$, and then  
    analyze the series 
for any given $U_0$.

We employ the following updates based on a slight modification 
    of the scheme used in 
    Ref.~\cite{polaron_lecture_notes,VANHOUCKE201095} to calculate the isospin susceptibilities within ladder series by calculating separtely the particle-hole polarization bubble with the dressed  four-fermion vertex dressed,   and the interacting part of the free energy (Luttinger-Ward functional) separately with the only difference that the susceptibilities contain isospin vertices.
\begin{enumerate}
\item Change $\tau$: This update is implemented only 
for the lowest-order contribution to susceptibility 
when $\mathcal{D}$ is the product of $G(\textbf{k},\tau)$ and $G(\textbf{k},-\tau)$. 
%
%
    Since exponential factors in Eq.~\ref{eq:Green's function} cancel out, this update is always accepted.
The lowest-order LW diagrams are only Hartree and Fock diagrams, both of which only involve Green's function at $\tau=0$. 
%
\item Change $\textbf{k}$: This update is also implemented only at the lowest order. For susceptibility, we change the momenta of both Green's functions from $\textbf{k}$ to $\textbf{k}'$, and the acceptance probability is
\begin{equation}
    P=\left|\frac{n_F(\epsilon_{\textbf{k}',\sigma,\gamma})(1-n_F(\epsilon_{\textbf{k}',\sigma,\gamma}))}{n_F(\epsilon_{\textbf{k},\sigma,\gamma})(1-n_F(\epsilon_{\textbf{k},\sigma,\gamma}))}\right|.
\end{equation}
For the LW functional, we only change 
   momentum for one $G$.
For the Hartree diagram, 
the acceptance probability is
\begin{equation}
    P=\left|\frac{n_F(\epsilon_{\textbf{k}',\sigma,\gamma}))}{n_F(\epsilon_{\textbf{k},\sigma,\gamma})}\right|.
    \label{PHartree}
\end{equation}
   For the Fock diagram, the update is rejected if the momentum transfer after the update is larger than $q^*$.
{Otherwise, the acceptance probability is given by
(\ref{PHartree}).
}
\item 
Swap valley: This update is necessary for two-valley systems with elliptic dispersion and is allowed only in the bare susceptibility and lowest order Hartree diagram of Luttinger-Ward functional. This is similar to changing $\mathbf{k}$ except here we change the valley index from $\gamma$ to $-\gamma$ and substitute the same $\mathbf{k}$ into the dispersion of the other valley. For the bare susceptibility the acceptance probability is 
\begin{equation}
    P=\left|\frac{n_F(\epsilon_{\textbf{k},\sigma,-\gamma})(1-n_F(\epsilon_{\textbf{k},\sigma,-\gamma}))}{n_F(\epsilon_{\textbf{k},\sigma,\gamma})(1-n_F(\epsilon_{\textbf{k},\sigma,\gamma}))}\right|.
\end{equation}
In the Hartree diagram, we choose one of the two Hartree bubbles at random, and the acceptance probability is
\begin{equation}
    P=\left|\frac{n_F(\epsilon_{\textbf{k},\sigma,-\gamma}))}{n_F(\epsilon_{\textbf{k},\sigma,\gamma})}\right|.
    \label{PHartree valley}
\end{equation}

\item Add Hartree: 
 This update allows us to change the diagram order, although diagrams with any Hartree bubbles are not included in the measurements since the renormalization of $G$ is not considered.
We randomly choose a Green's function in the present configuration, an imaginary time 
    $\tau \in (0, \beta=1/T)$, 
    and a momentum $\textbf{k}$ from a circle 
    of radius $1$, which is the high-momentum cutoff we impose on the dispersion. 
    In two-valley systems with elliptic dispersion in the states where valley is unpolarized, we also select the valley indices $\gamma$ with probability $1/2$. 
    If the selected Green's function parameters are
    $\textbf{p}$, $\tau_b$ and $\tau_a$,  then the acceptance probability is given by
\begin{equation}
    P=\left|\frac{f\beta N_{GF}\pi U_0G_{\sigma,\gamma}({\textbf{k}},0)G_{\sigma,\gamma}({\textbf{p}},\tau-\tau_b)G_{\sigma,\gamma}(\textbf{p},\tau_a-\tau)}{4\pi^2(N_{\textrm{Hartree}}+1)G_{\sigma,\gamma}(\textbf{p},\tau_a-\tau_b)}\right|,
\end{equation}
where $f$ is the number of occupied Fermion isospins with the same dispersion (For one-valley systems, $f=2$ in the unpolarized state and $f=1$ in the fully polarized ferromagnetic state. For two-valley systems with isotropic dispersion, $f=4$ in the unpolarized  state, and $f=1$ in the fSVP state. For two-valley systems with elliptic dispersion, $f=2$ in the unpolarized and fVP states, and $f=1$ in fSVP and fFM states),
$N_{GF}$ is the total number of Green's functions, and $N_{\textrm{Hartree}}$ is the number of Hartree bubbles
prior to the update. Care should be taken if the selected Green's function already belongs to a Hartree bubble.
In such a case, the $+1$ term in the denominator should be removed since $N_{\textrm{Hartree}}$ does not change after the update (see Fig.~\ref{fig:hartree on hartree}). 
 
 \item Remove Hartree: 
   This update removes a Hartree bubble selected at random
   from the diagram. This update decreases the diagram by one and is complementary to the "Add Hartree" update.
   See Fig.~\ref{fig:addremove} for illustration of both updates. 
In the same notation, the acceptance probability is given by 
\begin{equation}
P=\left|\frac{4\pi^2N_{\textrm{Hartree}}G_{\sigma,\gamma}({\textbf{p}},\tau_a-\tau_b)}{f\beta\pi(N_{GF}-2)G_{\sigma,\gamma}({\textbf{k}},0)G_{\sigma,\gamma}({\textbf{p}},\tau-\tau_b)G_{\sigma,\gamma}(\textbf{p},\tau_a-\tau)}\right|.
\end{equation}

\item  Swap $U$: 
     The proposal is to swap end points of two interaction lines connected by a Green's function (this description is topological; 
     interaction lines do not change their variables in the update, only
     Green's functions are modified)
   
     With the imaginary time and momenta specified in Fig.~\ref{fig:update} the acceptance probability is given by 
\begin{equation}
    P=\left|\frac{G_{\sigma,\gamma}({\textbf{k}},\tau_k-\tau_i)G_{\sigma,\gamma}({\textbf{k}+\textbf{p}-\textbf{q}},\tau_j-\tau_k)G_{\sigma,\gamma}({\textbf{q}},\tau_l-\tau_j)}{G_{\sigma,\gamma}({\textbf{k}},\tau_j-\tau_i)G_{\sigma,\gamma}({\textbf{p}},\tau_k-\tau_j)G_{\sigma,\gamma}({\textbf{q}},\tau_l-\tau_k)}\right|.
\end{equation}
\item Swap $G$: This update changes the number of fermion loops. 
We randomly choose an interaction line 
and propose to swap the destination points for two outgoing Green's functions. Since the values of the Green's functions remain the same, 
the acceptance probability depends only on the momentum transfer 
$Q=|\mathbf{p}-\mathbf{q}-\mathbf{k}|$ and diagram topology. 
If $Q<q^*$ the update is automatically accepted if the outgoing Green's functions belong to the same fermion loop and, thus, have the same isospin index, or it is accepted with probability $P=1/f$ if these Green's functions belong to different Fermion loops (the diagram weight is proportional to $f^\ell$ where 
$\ell$ is the number of fermionic loops).
Naively one would assume the update should be automatically rejected at $Q>q^*$, e.g. exchange interaction between two valleys (see Fig.~\ref{fig:exchange}).
\begin{figure}
    \centering
    \includegraphics[width=\linewidth]{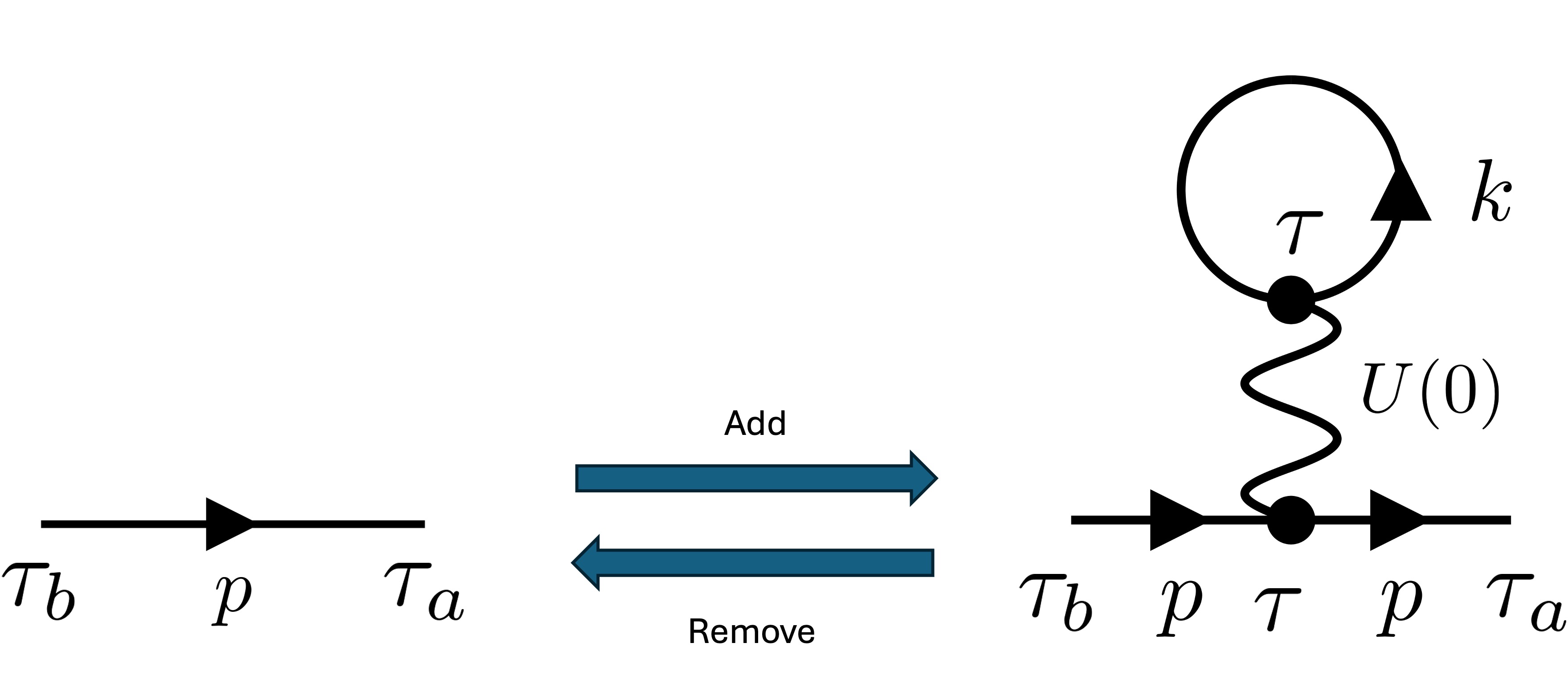}
    \caption{Illustration of the Add Hartree/Remove Hartree updates.}
    \label{fig:addremove}
\end{figure}
\begin{figure}
    \centering
    \includegraphics[width=\linewidth]{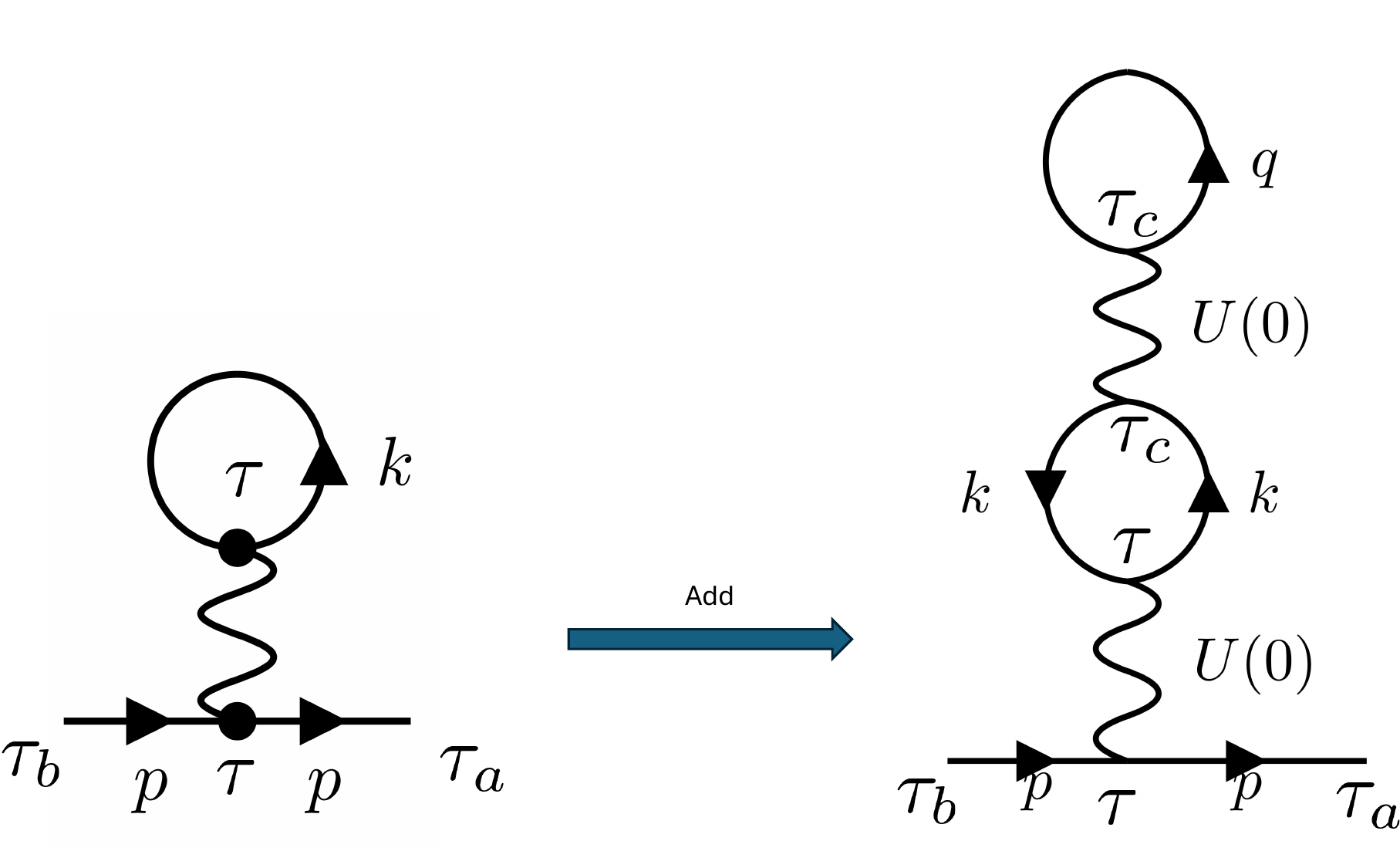}
    \caption{Demonstration of adding  a Hartree bubble on a Green's function which is a part of a Hartree bubble eliminates the existing one. As one can see the number of Hartree bubbles remain the same before and after the update.}
    \label{fig:hartree on hartree}
\end{figure}
\begin{figure}
    \centering
    \includegraphics[width=\linewidth]{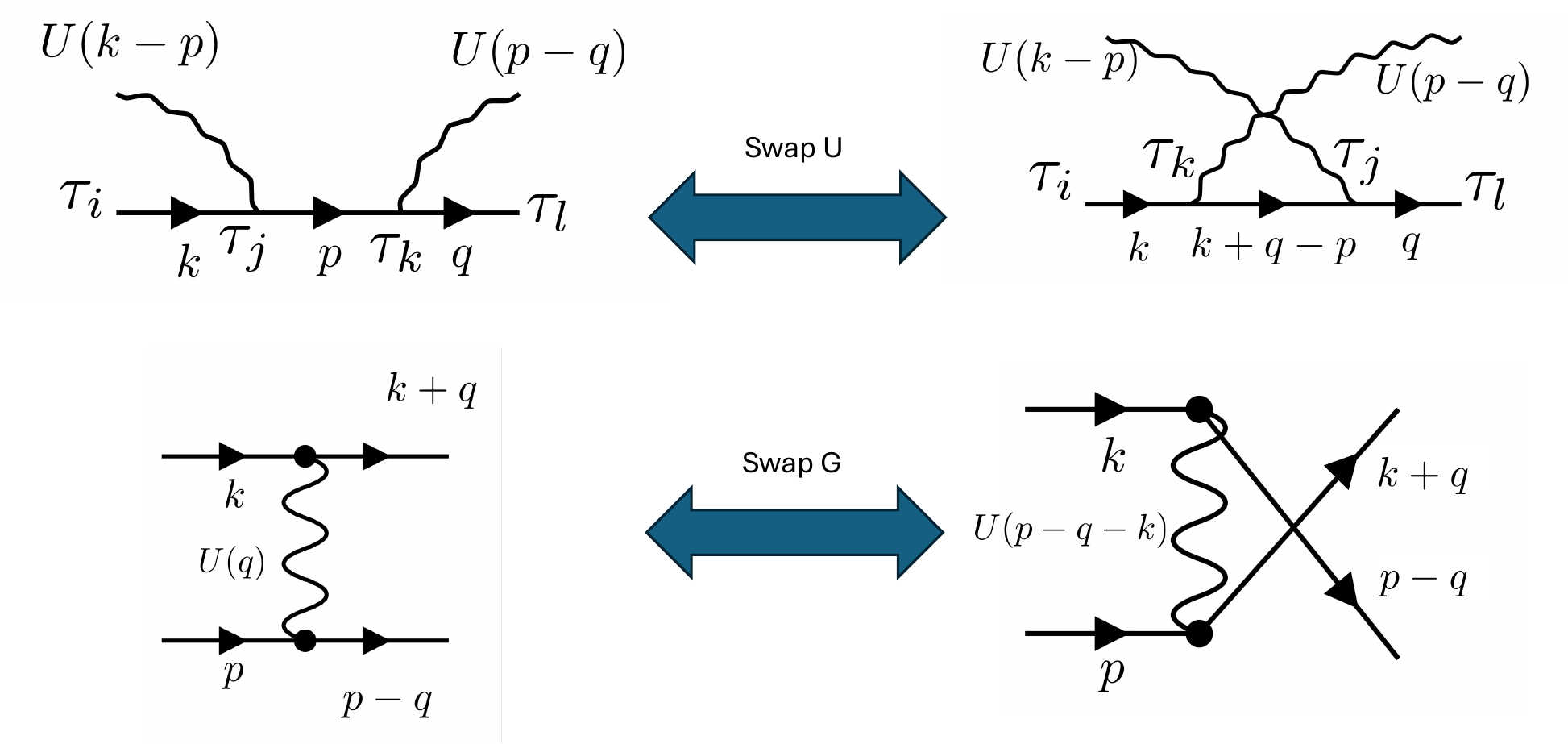}
    \caption{Illustration of updates that swap the interaction lines (top panel) and the Green's functions (bottom panel).}
    \label{fig:update}
\end{figure}
However, for ergodicity, we shall accept it with arbitrary probability to ensure all the diagrams connected by density-density interactions are sampled despite not including the exchange diagrams from the measurement. In Fig.~\ref{fig:example} we give an example of obtaining one of the diagrams in LW functional in a two-valley system with elliptic dispersion to second order in $U_0$ via a series of updates starting from the lowest order Fock diagram, which is made possible only by allowing the unphysical intermediate diagrams that include intervalley exchange interaction. The same argument also holds for one-valley systems when $Q>q^*$.
\begin{figure}
    \centering
    \includegraphics[width=\linewidth]{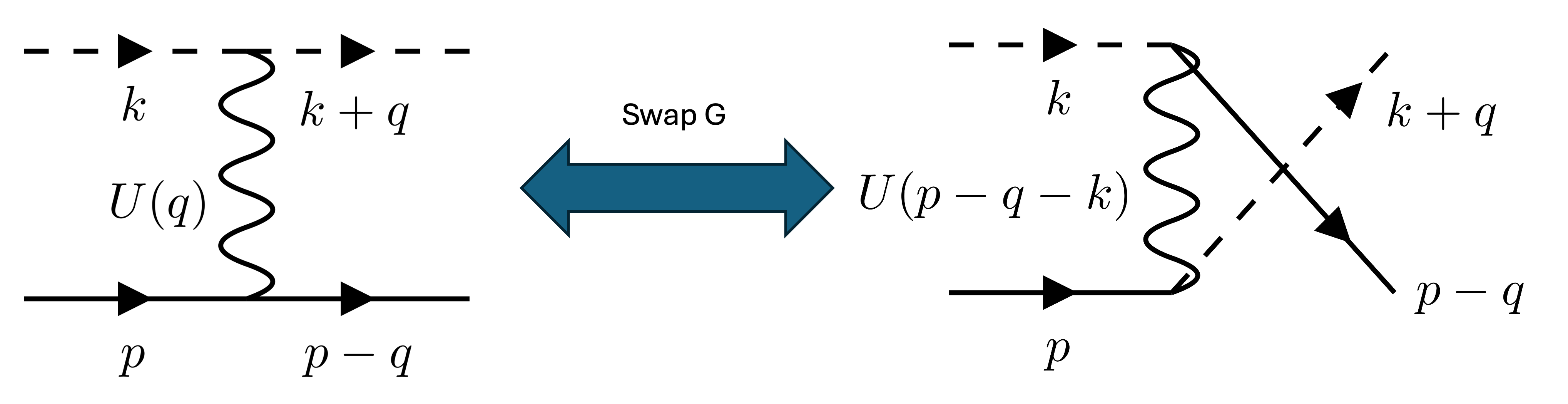}
    \caption{The Swap G update that results in intervalley exchange interaction. Here two distinct valleys are labeled by solid and dashed lines, respectively. The momentum $k$ and $p$ are in the vicinity of the two valley centers and $q$ is assumed to be much smaller than the large momentum cutoff $1$, so the momentum transfer after the update $p-q-k$ is roughly the distance between the valleys $\pi/2$.}
    \label{fig:exchange}
\end{figure}
\begin{figure}
    \centering
    \includegraphics[width=\linewidth]{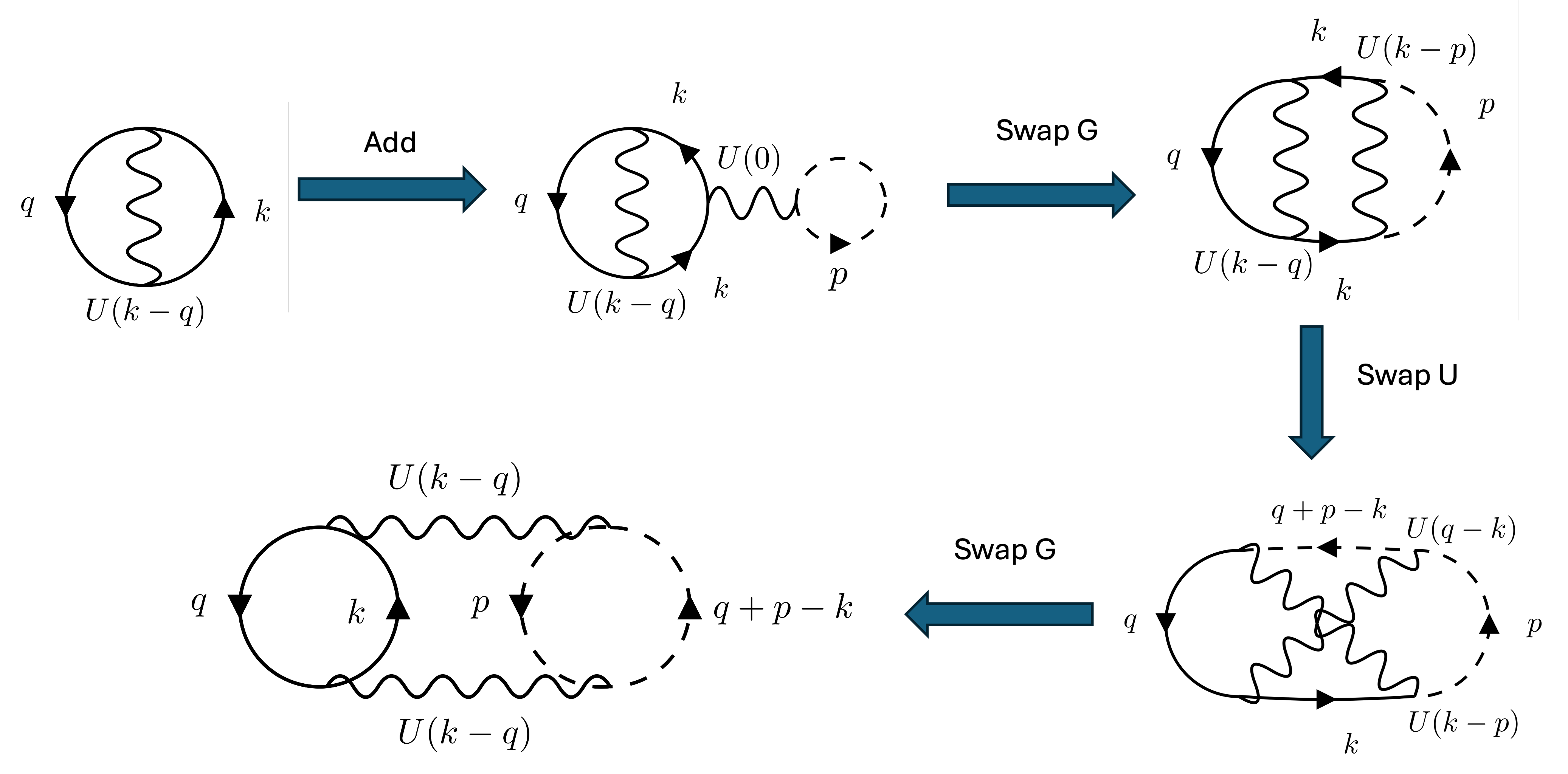}
    \caption{An example of obtaining a term in Luttinger-Ward functional in a two-valley system with elliptic dispersion to first order in $U_0$ under a sequence of updates. The intermediate diagrams include intervalley exchange interaction, which is large in momentum transfer and therefore not included in the measurement. Here solid and dashed arrows denote the Green's functions in two different valleys, i.e. $k$ and $q$ are in the vicinity of the valley $\gamma=1$ and $p$ is in the vicinity of $\gamma=-1$.}
    \label{fig:example}
\end{figure}
Through this update scheme, the diagram is guaranteed to be connected. As mentioned in the main text,  we only include the two-particle irreducible diagrams in the measurement. To exclude the diagrams which do not fall into this category, we do not include the diagrams with Green's functions of repeating momentum labels in the measurements, except the two Green's function connected by the isospin vertices when calculating susceptibilities since we are considering $\mathbf{q}=0$ order. With this procedure, electron self-energy corrections, including Hartree and Fock, are also excluded, since the momentum before and after Hartree/Fock are the same.
\end{enumerate}
\subsection{Resummation scheme}\label{app:resummation}
\begin{figure}
    \centering
    \includegraphics[width=\linewidth]{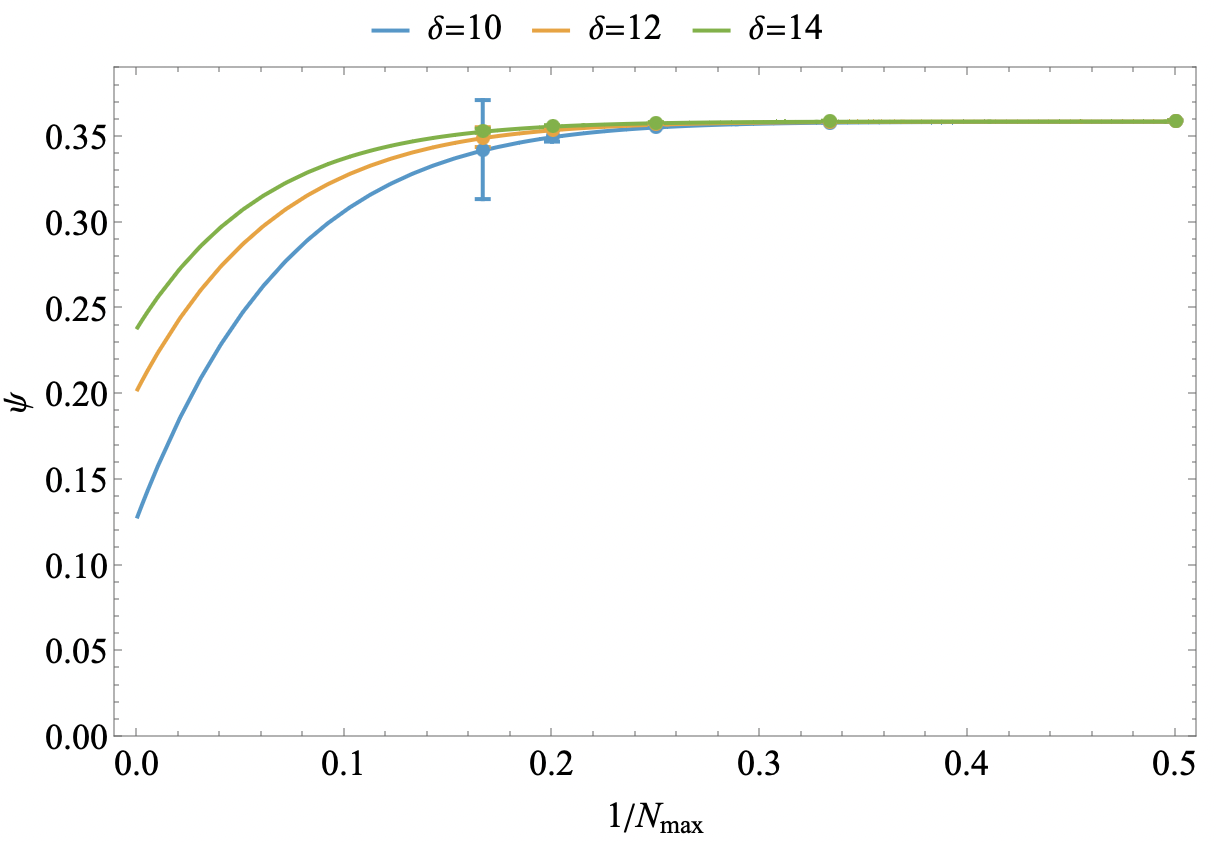}
    \caption{The ratio $\psi$ defined in Eq.~\ref{eq:psi} for $q^*=k_F$ at $\lambda=1$  obtained by extrapolation using different exponents in Fig.~\ref{eq:Cesaro}. }
    \label{fig:extrapolate}
\end{figure}

      In practice, one can only compute expansion coefficients up to a finite order, $N_{max}$. The series will oscillate with ever-increasing amplitude when they happen to be outside of the radius of convergence, in which case the extrapolation of results to an infinite order is achieved by applying resummation techniques. 
      One procedure used in this work is to
multiply each term $\mathcal{D}_n$ by a factor $g(n,N_{max})$ such that $g$ goes to one when $N_{max}\rightarrow\infty$ and $g$ decays faster than an exponential function when $n\rightarrow\infty$. With that, we can instead calculate the series
\begin{equation}
    \mathcal{O}(N_{max})=\sum_{n=0}^{N_{max}}\mathcal{D}_ng(n,N_{max}),
\end{equation}
plot $\mathcal{O}(N_{max})$ as a function of $1/N_{max}$, and get the result for $\mathcal{O}$ by extrapolating to $1/N_{max}\rightarrow 0$. The choice of $g$-function is arbitrary as long as it satisfies the above conditions.
To be concrete, we use the Riesz-Ces\`aro method with 
\begin{equation}
    g(n,N_{max})=[(N_{max}-n+1)/N_{max}]^\delta
    \label{eq:Cesaro}
\end{equation}
 with  {$\delta=10,12,14$}. In Fig.~\ref{fig:extrapolate}, we present the results using different exponents. The extrapolation error is given by the difference between the maximum and minimum of the curves evaluated at $1/N_{max}=0$.
\bibliography{biblio}
\end{document}